\documentclass[preprint]{aastex}
\usepackage{emulateapj5}
\input psfig.tex

\def\hst{{\it HST}}
\def\etal{\emph{et al.}\ }

\def\kms{km s$^{-1}$}
\def\apj{ApJ}
\def\aj{AJ}
\def\mnras{MNRAS}
\def\apjs{ApJS}

\def\aa{{A\&A}}
\def\aas{{ A\&AS}}
\def\aj{{AJ}}

\def\apj{{ApJ}}
\def\apjs{{ApJS}}
\def\asec{$^{\prime\prime}$}

\def\deg{$^{\circ}$}

\def\etal{{et al. }}

\def\micron{{$\mu$m}}
\def\mnras{{MNRAS}}
\def\nat{{Nature}}

\def\percm2{cm$^{-2}$}

\def\lax{{$\mathrel{\hbox{\rlap{\hbox{\lower4pt\hbox{$\sim$}}}\hbox{$<$}}}$}}
\def\gax{{$\mathrel{\hbox{\rlap{\hbox{\lower4pt\hbox{$\sim$}}}\hbox{$>$}}}$}}

\def\hii{\ion{H}{2}}

\slugcomment{May 10, 2001; accepted by {\it The Astronomical Journal.}}
\lefthead{Ravindranath et al.}
\righthead{Central Structural Parameters of Early-Type Galaxies}
\begin{document}

\title{Central Structural Parameters of Early-Type Galaxies as Viewed with 
{\it HST}/NICMOS\footnotemark[1]}

\footnotetext[1]{Based on observations made with the {\it Hubble Space 
Telescope}, which is operated by AURA, Inc., under NASA contract NAS5-26555.}

\author{
Swara Ravindranath\altaffilmark{2,3},
Luis C. Ho\altaffilmark{2},
Chien Y. Peng\altaffilmark{4}, 
Alexei V. Filippenko\altaffilmark{3}, \\ 
and Wallace L. W. Sargent\altaffilmark{5}
}

\altaffiltext{2}{The Observatories of the Carnegie Institution of Washington, 
813 Santa Barbara St., Pasadena, CA 91101-1292.}

\altaffiltext{3}{Department of Astronomy, University of California, 
Berkeley, CA 94720-3411.}

\altaffiltext{4}{Steward Observatory, University of Arizona, Tucson, AZ 85721.}

\altaffiltext{5}{Palomar Observatory, 105-24 Caltech, Pasadena, CA 91125.}

\setcounter{footnote}{5}

\begin{abstract}

We present surface photometry for the central regions of a sample of 33 
early-type (E, S0, and S0/a) galaxies observed at 1.6~$\mu$m ($H$ band) using 
the {\it Hubble Space Telescope (HST)}. Dust absorption has less of an impact 
on the galaxy morphologies in the near-infrared than found in previous work 
based on observations at optical wavelengths.  When present, dust seems to be 
most commonly associated with optical line emission.  We employ a new 
technique of two-dimensional fitting to extract quantitative parameters for 
the bulge light distribution and nuclear point sources, taking into 
consideration the effects of the point-spread function.  Parameterizing the 
bulge profile with a ``Nuker'' law (Lauer et al. 1995), we confirm that the 
central surface-brightness distributions largely fall into two categories, 
each of which correlates with the global properties of the galaxies.  ``Core'' 
galaxies tend to be luminous ellipticals with boxy or pure elliptical isophotes, 
whereas ``power-law'' galaxies are preferentially lower luminosity systems 
with disky isophotes.  The infrared surface-brightness profiles are very 
similar to the optical, with notable exceptions being very dusty objects.
Similar to the study of Faber et al. (1997) based on optical data, we find 
that galaxy cores obey a set of fundamental-plane relations wherein more 
luminous galaxies with higher central stellar velocity dispersions generally 
possess larger cores with lower surface brightnesses.  Unlike most previous 
studies, however, we do not find a clear gap in the distribution of 
inner cusp slopes; several objects have inner cusp slopes ($0.3 < \gamma 
< 0.5$) which straddle the regimes conventionally defined for core and 
power-law type galaxies.  The nature of these intermediate objects is 
unclear.  We draw attention to two objects in the sample which appear to be 
promising cases of galaxies with isothermal cores that are not the brightest 
members of a cluster.  Unresolved nuclear point sources are found in 
$\sim$50\% of the sample galaxies, roughly independent of profile type, with 
magnitudes in the range $m^{\rm nuc}_H$ = 12.8 to 17.4 mag, which correspond 
to $M_H^{\rm nuc}$ = $-$12.8 to $-$18.4 mag.  Although the detection rate of 
compact nuclei seems favored toward galaxies spectroscopically classified as 
weak active galactic nuclei, we find no significant correlation between the 
near-infrared nuclear luminosities and either the optical emission-line 
luminosities or the inferred black-hole masses.
\end{abstract}


\section{Introduction}                   %%1

Prior to the advent of the {\it Hubble Space Telescope (HST)}\, ground-based 
studies of luminous elliptical galaxies showed that the surface-brightness 
profiles had a central core in most cases, but the effect of atmospheric 
seeing (typically \gax 1\asec) made it difficult to discriminate between truly 
resolved isothermal cores and unresolved cores (Schweizer 1981; Kormendy 
1985a).  Kormendy (1985a, 1985b) studied elliptical galaxies and spiral bulges 
using images taken under excellent seeing conditions (0\farcs 2--0\farcs 5) 
and confirmed that isothermal cores were indeed very rare. However, some 
ellipticals showed evidence for isothermal cores, and these were generally the 
brightest galaxies in rich clusters.  Attempts were made to relate the core 
properties with the global properties by accounting for the effects of seeing 
through image deconvolution (Lauer 1985a).  However, resolution approaching 
0\farcs 1 would be required to resolve cores in low-luminosity galaxies for 
which the correlation between luminosity and core radius implies small core 
sizes (Kormendy 1985b). 

Results from $V$-band imaging of early-type galaxies using the Wide Field 
Planetary Camera (WFPC) showed that traditional functions used to fit 
ground-based surface-brightness distributions, such as King (1966) or 
de~Vaucouleurs (1948) $r^{1/4}$ profiles, do not provide adequate fits for the 
central ($r$ \lax 1\asec) regions (Crane et al. 1993; Ferrarese et al. 1994; 
Forbes, Franx, \& Illingworth 1995; Lauer et al. 1995).  The King models 
used for giant ellipticals have central cores with constant luminosity 
densities, which cause the brightness profiles to appear flat in the center.  
However, the {\it HST}\ studies argued against the existence of such 
isothermal cores based on the non-zero cusp slopes seen even within 
$r$ \lax 1\asec.  It was evident that some galaxies have profiles that can 
be represented by a single power law all the way to the \hst\ resolution 
limit, while others require double power laws, with the inner slope interior 
to some characteristic radius being much shallower than the outer slope.  This 
led to the formulation of an empirical function,  essentially a double power 
law, to describe the surface-brightness profiles.  The ``Nuker'' law (Lauer et 
al. 1995; Byun et al. 1996) has the functional form

\begin{equation}
I(r)=2^{(\beta-\gamma)/\alpha}I_{b}\left(\frac{r}{r_{b}}\right)^{-\gamma}
\left[1+\left(\frac{r} {r_{b}}\right)^{\alpha}\right]^{(\gamma-\beta)/\alpha},
\end{equation}

\noindent
where $\beta$ is the asymptotic slope as $r \rightarrow \infty$, $\gamma$
is the asymptotic slope as $r \rightarrow 0$, $r_{b}$ is the break radius at 
which the outer slope $\beta$ changes to the inner slope $\gamma$, $\alpha$ 
controls the sharpness of the transition between the inner and outer slopes, 
and $I_{b}$ is the surface brightness at $r_{b}$.  

Two classes of early-type galaxies can be identified based on the value of 
$\gamma$ (Lauer et al. 1995).  ``Core'' galaxies have surface-brightness 
profiles which exhibit a significant flattening within a well resolved 
$r_{b}$ (\gax 0\farcs2), such that $\gamma$ \lax 0.3, whereas the profiles of 
``power-law'' galaxies do not show any significant break but continue to be 
steep ($\gamma$ \gax 0.5) up to the resolution limit of $r\,\approx$ 0\farcs1.
There was an apparent dichotomy in the distribution of $\gamma$ values.  Core 
galaxies have $0 <  \gamma \leq 0.3$, and power-law galaxies show 
$\gamma \geq 0.5$; none seemed to have $\gamma$ in the range 0.3 to 0.5 (Faber 
et al. 1997).  Core profiles are found predominantly in luminous, slowly 
rotating systems with boxy or pure elliptical isophotes, while power-law 
profiles occupy less luminous, rapidly rotating galaxies with disky isophotes 
(Jaffe et al. 1994; Byun et al. 1996; Faber et al. 1997).  
Carollo et al. (1997) obtained Wide 
Field Planetary Camera 2 (WFPC2) $V$-band and $I$-band images for a sample of 
elliptical galaxies with kinematically distinct cores. Parameterizing the 
surface-brightness profile using the Nuker law, they too found that 
fast-rotating, disky galaxies have steep inner slopes while slow-rotating, 
boxy systems tend to have shallow inner slopes.  In spite of these trends with 
global galaxy properties, which are similar to those of kinematically normal 
ellipticals, they suggest that the inner slopes may vary continuously between 
core and power-law galaxies.  Recently, a detailed analysis of the 
surface-brightness profiles for early-type galaxies has been carried out by 
Rest et al. (2001) using $R$-band WFPC2 images. Their results, also based 
on Nuker fits to the profiles, present further evidence for a continuous 
distribution of inner slopes. 

This paper analyzes $H$-band (1.6 \micron) images of a sample of 33 early-type 
(E, S0 and S0/a) galaxies observed with NICMOS on {\it HST}.  Our primary aim 
is to parameterize the intrinsic distribution of the bulge light.  We achieve
this using a two-dimensional (2-D) decomposition technique which deblends 
the bulge from a central nucleus and properly accounts for the NICMOS 
point-spread function (PSF).  As is well known from the WFPC and WFPC2 
studies, even early-type galaxies contain considerable amounts of dust in 
their centers which corrupts the surface-brightness profiles and isophotal 
parameters (van~Dokkum \& Franx 1995; Verdoes Kleijn et al. 1999; Tomita et 
al. 2000; Tran et al. 2001).  Near-infrared (NIR) images are better suited for 
isophotal analysis because of the reduced sensitivity to dust extinction and 
emission from young star clusters.  They also trace more faithfully the 
underlying stellar population dominating the mass.  

The paper is organized as follows.  Section 2 describes the sample and 
data reduction.  Section 3 summarizes our methods of isophotal analysis, 
including an introduction of our 2-D modeling technique.  The principal 
results concern small-scale structures in the inner regions (\S~4), 
surface-brightness profiles and central-parameter relations (\S~5), and
unresolved nuclei (\S~6).  We discuss the implications of our findings 
in \S~7 and present a summary in \S~8.  Appendix A gives noteworthy
details on individual objects.   Throughout this paper distance-dependent 
quantities are based on a Hubble constant of $H_0$ = 75 \kms\ Mpc$^{-1}$.

\section{The Sample and Data Reduction}        %%2
%
%%%%%%%%%%%%%%%%%%%%%%%%%%%%%%%%%%%%%%%%%%%%%%%%%%%%%%%%%%%%%%%%%%%%%%%%%%%
%
%%good bounding box = 20 90 600 720
%\begin{figure*}[t]
%\centerline{\psfig{file=table1_rev.ps,width=17.5cm,angle=0}}
%\end{figure*}
%
%%%%%%%%%%%%%%%%%%%%%%%%%%%%%%%%%%%%%%%%%%%%%%%%%%%%%%%%%%%%%%%%%%%%%%%%%%
%
%%%%%%%%%%%%%%%%%%%%%%%%%%%%%%%%%%%%%%%%%%%%%%%%%%%%%%%%%%%%%%%%%%%%%%%%%%
%
%%good bounding box = 20 90 600 720
%\begin{figure*}[t]
%\centerline{\psfig{file=table2_rev.ps,width=17.5cm,angle=0}}
%\end{figure*}
%
%%%%%%%%%%%%%%%%%%%%%%%%%%%%%%%%%%%%%%%%%%%%%%%%%%%%%%%%%%%%%%%%%%%%%%%%%%
%
Our sample is based on galaxies selected from the Palomar study of nearby 
galaxies, a ground-based optical spectroscopic survey of a nearly complete 
sample of 486 galaxies with $B_T\,\leq$ 12.5 mag and declinations greater than 
0\deg\ (Filippenko \& Sargent 1985; Ho, Filippenko, \& Sargent 1995, 1997a, 
1997b).  Our study focuses on the subset of early-type galaxies with NICMOS 
images available in the {\it HST}\ data archive.  This comprises 33 objects 
(14 E, 16 S0, and three S0/a) from different snapshot programs.  The global 
properties of the sample galaxies are given in Table~1, along with their 
nuclear spectral classifications, which includes five Seyferts, 15 
low-ionization nuclear emission-line regions (LINERs), three ``transition'' 
(LINER/\hii) objects, one \ion{H}{2} nucleus, and nine absorption-line nuclei 
(see Ho et al. 1997a for a description of the classification system).  

The data used in this work consist of images obtained in the F160W filter 
($H$ band) using the NIC2 and NIC3 cameras. The NIC2 images have a field of 
view of 19$\farcs$2 $\times$ 19$\farcs$2 and a pixel scale of 0$\farcs$076;
the gain and read noise are 5.4 e$^{-}$ count$^{-1}$ and 30 e$^{-}$, 
respectively. The NIC3 camera provides a field of view of 
51$\farcs$2 $\times$ 51$\farcs$2 and a pixel scale of 0$\farcs$2; its 
gain and read noise are 6.5 e$^{-}$ count$^{-1}$ and 30 e$^{-}$, respectively. 
The majority of the data were taken as part of snapshot survey programs, 
and therefore the exposure times were relatively short, ranging from 15 to 
320~s, typically $\sim$150~s.

Data reduction begins with the standard (``calnicA'') pipeline processing 
performed at the Space Telescope Science Institute (STScI; Bushouse et al. 
1998). The calnicA task works on the raw science data and removes the 
instrumental signatures through bias subtraction, dark-current correction, 
nonlinearity correction, and flat fielding. A ramp-fitting procedure removes 
most of the cosmic rays from the images. A few additional processing steps, as 
recommended in the NICMOS Handbook (Dickinson 1999), are required to remove 
residual image anomalies.  The NICMOS arrays have a number of bad pixels, some 
of which have very low response and appear dark (cold pixels) and others which 
have very high or erratic dark current and appear bright (hot pixels) in the 
images. In addition to bad pixels, there are regions of reduced sensitivity, 
termed ``grots,'' which result from flecks of anti-reflective paint that have 
scraped off from the optical baffles. The best way to eliminate the effects of 
bad pixels would be to use dithered images. Since the archival images we use 
originate from various observing programs, dithered images are lacking for 
some galaxies, and in a few cases only single exposures are available.  
Therefore, we created a mask of all bad pixels from the flat-field images 
using the {\it ccdmask}\ task in IRAF\footnote{IRAF is distributed by the 
National Optical Astronomy Observatories, which are operated by the 
Association of Universities for Research in Astronomy, Inc., under cooperative 
agreement with the National Science Foundation.} and then used the 
{\it fixpix}\ task to interpolate over the masked pixels.

The NIC2 camera has a coronographic hole which is used when observing 
targets close to bright objects. In the non-coronographic F160W observations 
used for this study, the hole appears in the images as a bright circular 
patch with positive intensity due to excess background emission from warmer 
structures located behind it.  The position of the coronographic hole 
with respect to the detector coordinates is known to have moved with time, 
and this causes two patches to appear in the calibrated image, the second 
one arising from the use of reference files taken at a different time. We 
masked the coronographic hole during further analysis. 

The calnicA-processed images contain a residual offset or ``pedestal'' which 
appears as a time-variable bias that varies from one quadrant to another. This 
variable quadrant bias is believed to result from subtle temperature changes 
in the electronics or the detectors themselves. The four quadrants of each 
NICMOS detector have separate amplifiers, thereby resulting in a variation of 
the pedestal among them. The effect of the residual bias, which ranges from 
0.1 to 0.35 count s$^{-1}$, becomes evident when the dark-subtracted images 
are multiplied by the flat-field reference files, whereby the relative pixel 
sensitivities dominate the images (B\"{o}ker et al. 1999).  We used software 
developed by R.~van~der~Marel at STScI to remove the pedestal 
effect\footnote{http://sol.stsci.edu/\~marel/software/pedestal.html}.
 
An important issue when seeking information at the highest spatial resolution 
is to understand the properties of the PSF.  The PSF often defines the 
resolution and sensitivity limits of the observation and can change with time, 
wavelength, position, and the camera used. At 1.6~$\mu$m, the PSF is 
critically sampled for the NIC2 camera (core FWHM $\approx$ 2.3 pixels, or 
0\farcs17), while it is undersampled for the NIC3 camera (core FWHM $\approx$ 
1.1 pixels, or 0\farcs22). In NIC2 images the drift in the cold masks with 
time causes variations in the spider patterns and diffraction rings of the 
PSF. However, the PSFs of all the NICMOS cameras can be well modeled by the 
Tiny Tim software (Krist \& Hook 1999), which takes these effects into 
account. In the absence of PSFs empirically derived from observations of 
bright stars, synthetic PSFs generated by Tiny Tim serve as good alternatives 
for use in photometry, deconvolution, and image modeling (Krist \& Hook 1997).  

\vskip 0.3cm
\section{Surface Photometry}              %%3

\subsection{Isophotal Analysis}

Galaxies appear relatively smooth in NICMOS images compared to the optical 
images, and thus are well suited for deriving surface-brightness profiles.
We used the {\it ellipse}\ task in STSDAS to perform surface photometry. This 
task assumes that the isophotes of a galaxy can be represented by ellipses. 
The ellipse fit is performed by providing as input parameters an initial guess 
value for the galaxy centroid, the ellipticity, and the position angle, and 
then allowing all the parameters to vary with increasing semi-major axis 
radius. The initial coordinates for the galaxy center are estimated using the 
{\it imexamine}\ task.  However, the isophotes of a galaxy are often not 
perfect ellipses. The deviation from the fitted ellipses is quantified by the 
higher-order coefficients in the Fourier series

\begin{equation}
I(\phi)=I_{o}+\sum_{n=1}^{N}\left [A_{n}{\rm sin}(n\phi)+B_{n}{\rm cos}(n\phi)\right] ,
\end{equation} 

\noindent
where $N$ is the highest order fitted and $\phi$ is the angle measured 
counter-clockwise from the major axis of the ellipse.  The amplitudes of the 
second-order Fourier terms ($A_2$ and $B_2$) are used to obtain the position 
angle (P.A.) and ellipticity ($\epsilon$) of the best-fitting ellipse. A 
perfectly elliptical isophote is completely described by the first and 
second-order Fourier terms of the above equation. Non-zero coefficients for 
the higher-order terms in the expansion series carry information on the shape 
of the isophotes.  These intensity coefficients are divided by the semi-major 
axis length and local intensity gradient to measure actual radial deviations 
from perfect ellipses (Jedrzejewski 1987). The third-order terms have been 
found to have significant non-zero values when the isophotes get skewed by the 
presence of dust (Peletier et al. 1990). The most dominant mode which carries 
information about the isophote shapes is the fourth-order cosine term,
and its amplitude $B_4$ is positive for disky isophotes and negative for boxy 
isophotes. A non-zero value for the corresponding sine term indicates rotation 
or misalignment relative to the major axis of the ellipse which could result 
from the presence of dust or projection effects (Franx, Illingworth, \& 
Heckman 1989).

The results of the isophotal analysis are shown in Figure~1. Note that our 
ellipse fits are performed on the {\it observed} (without deconvolution) 
images.  (This is in contrast to the 2-D fits in \S~3.2, where we properly 
account for the PSF.) The effects of the PSF can be seen in the lower three 
panels of Figure~1, where interior to $r$ \lax 0\farcs2--0\farcs3 the values of 
$\epsilon$, P.A., and $B_4$ have large formal error bars and often undergo 
erratic variations. The large fluctuations at small radii are also partly 
caused by the discrete sampling and sub-pixel interpolation implemented in 
the {\it ellipse}\ routine (see discussion in Rest et al. 2001).  We consider 
these data points to be unreliable.  The radial variations in ellipticity and 
P.A. at large radii are caused by the presence of features such as dust lanes, 
weak disks, or nuclear bars, and thus reveal morphological peculiarities in 
galaxies. In some cases, the variations in ellipticity and P.A. may also 
reflect triaxiality in galaxies.
The $H$-band magnitudes were computed using the photometric keywords provided 
in the image header for converting observed counts to the Vega magnitude 
system (or the Johnson $H$-band magnitude). The conversion is 
$m_{H}$ = --2.5 log(count s$^{-1}$) + $m_{\rm zpt}$, where $m_{\rm zpt}$ = 
21.75 for NIC2 and 21.50 for NIC3.  

\subsection{2-D Modeling}  %

Ground-based studies of the surface-brightness profiles of galaxies 
traditionally employ 1-D analysis, whereby the radial profile is first derived 
through isophote fitting, followed by decomposition with a combination of an 
exponential disk and a de~Vaucouleurs bulge profile (e.g., Freeman 1970; 
Kormendy 1977; Boroson 1981; Kent 1985; Baggett, Baggett, \& Anderson 1998).  
However, in recent years the importance of using the entire 2-D image to 
obtain more accurate surface photometric parameters has been extensively 
discussed in the literature (Byun \& Freeman 1995; de~Jong 1996; Moriondo, 
Giovanardi, \& Hunt 1998; Scorza et al. 1998; Wadadekar, Braxton, \& Kembhavi 
1999).  2-D image decomposition proves superior to decomposition in 1-D when 
the shape parameters are significant, for instance when a strong disk is 
present.

When decomposing galaxies into constituent components, modeling degeneracy is 
a serious issue, particularly in the context of bulge-to-disk decomposition 
(e.g., Kent 1985; Byun \& Freeman 1995; Moriondo et al. 1998). For example, 
a combination of the de~Vaucouleurs law and an exponential disk has been 
widely used, but they usually fit well only over a limited range in radius, 
and often poorly near the nucleus.  A certain subjectivity is sometimes 
involved, therefore, in deciding where and how to fit --- a decision which 
affects the derived bulge and disk parameters.  An additional complication 
is that there is no reason why the popular $r^{1/4}$ function should be 
preferred over, say, the more general formulation of S\'ersic (1968), which 
often may fit just as well, if not better.  These problems can be serious 
in 1-D modeling.  But 2-D modeling, which uses the full spatial information, 
offers the potential to break the degeneracies: bulges and disks can often 
be decoupled in 2-D because they may have different position angles and 
ellipticities.  The appropriateness of the chosen functions is also usually 
more apparent.

In a similar vein, significant ambiguity can arise when attempting to extract 
compact sources (nonstellar emission from active galactic nuclei or nuclear 
star clusters) in the centers of galaxies.  The degeneracy here is caused by
the seeing, which reduces the contrast between the nuclear source and the 
underlying galaxy profile, which itself can be sharply peaked.  Earlier 
studies of nearby nuclei extracted central sources by fitting 1-D galaxy 
profiles simultaneously with a point source (e.g., Carollo, Stiavelli, \& 
Mack 1998).  The results can depend sensitively on details such as the 
manner in which the 1-D profile is obtained.  Ferrarese et al. (1994) find 
that while there is no noticeable difference in the surface-brightness 
profiles along the major and minor axes for core galaxies, the same is not 
true for the power-law galaxies.  Contamination by dust can affect the centering 
and distort the overall nuclear profile, especially in undersampled images.  
But in 2-D, one can again use the shape information present in the entire 
image to one's advantage.  Wadadekar et al.  (1999) illustrate the benefit of 
using 2-D modeling over 1-D for extracting central point sources.

In the context of {\it HST}\ images, Lauer et al. (1998) showed that even with
WFPC2 images, PSF smearing significantly affects a galaxy's profile within 
0\farcs5.  To remove the effect one can either deconvolve the image or 
convolve a model profile with the PSF to match the observed images.  Which 
strategy to adopt depends on the signal-to-noise ratio of the data, and to a 
lesser extent on knowledge of the PSF.  Our NICMOS images come mainly from 
various snapshot survey programs and hence do not have uniformly high 
signal-to-noise ratios to enable reliable deconvolution.  We also do not have 
access to near-contemporaneous observations of stars to derive empirical 
PSFs.  The synthetic Tiny Tim PSFs are not known to high accuracy because of 
thermal stresses and ``breathing'' effects. We decided, therefore, that 
convolution is the most transparent way to proceed.

We parameterize the bulge light distribution of our sample of early-type
galaxies using a 2-D Nuker profile.  This was done using a least-squares 
fitting program, GALFIT (C.~Y. Peng et al., in preparation), which allows 
fitting a superposition of 
analytic models (e.g., S\'ersic/de~Vaucouleurs,
Nuker, exponential, Gaussian, Moffat) to create arbitrarily complex galaxy
shapes.  The radial shapes of the models are generalized ellipses
(Athanassoula et al. 1990), which have a radius given by

\begin {equation}
r^{c+2}=\left(\left|x\right|^{c+2} +
                 \left|{y\over q}\right|^{c+2}\right),
\end {equation}

\noindent
where $q$ is the axis ratio, and $c$, as defined, permits the ellipses to be
either disky ($c < 0$) or boxy ($c > 0$). The shape parameters of the generalized
ellipses ($c$, $q$, the center, and P.A.) are free parameters in the fit, but do
not vary as a function of radius. 
To optimize the fit, GALFIT uses a Levenberg-Marquardt algorithm 
(Press et al. 1992), one that traverses down a $\chi^2$ gradient toward a minimum. 
This algorithm is very fast compared to
random-walk type algorithms (e.g., Simulated Annealing or Metropolis) at
probing large parameter spaces, but has the tendency to ``fall'' into a local 
minimum and be content.  However, reasonable initial values and probing 
parameter space near a minimum can often overcome this shortcoming. 
Furthermore, in light of the fact that model functions do not perfectly fit 
galaxies, the globally minimum $\chi^2$ may not always produce the most 
meaningful fit.

In order to compare with past WFPC2 and NICMOS studies, we have decided to
use a single 2-D Nuker function to parameterize the galaxy, even if this choice
may not reproduce all the complex structures near the nucleus.  Better fits
can often be achieved with two or more subcomponents (L.~C.~Ho et al. 2001, 
in preparation).  We create a Nuker model (equation 1) over the dimension of 
the entire NICMOS image.  We then convolve it with a Tiny Tim PSF by following
the convolution theorem. The fast Fourier transforms of the model image and 
PSF image are multiplied in complex space and then inverse transformed to 
obtain the final convolved model image.
The generated model is a single bulge component described by the Nuker
law with optimized shape parameters that provide the best global fit. 
The Nuker parameters $\alpha$, $\beta$, and $\gamma$ are constrained to be 
positive. The effects of PSF undersampling in the NICMOS images can be very
significant, especially when the profile is cuspy near the nucleus. It is
important that the contributions from the cusp center and the surrounding
portions over the pixels are extracted correctly (Rest et al. 2001).
In GALFIT, the inner five pixels are divided into elliptical-polar grids
centered on the galaxy, with radial spacings that increase in sampling closer
and closer to the center. Thus, the integration is performed with increased
subsampling for the inner pixels.  

The necessity of including an additional component for a central compact 
source is obvious from the surface-brightness profiles in most cases, 
consistent with the fact that a large fraction of the sample is 
spectroscopically classified as active galactic nuclei (AGN).  
In these cases we added a ``point'' 
source on top of the galaxy.  The point source is approximated by a narrow 
Gaussian convolved with the PSF.  Using a Gaussian, we can determine if a 
central source is truly unresolved (intrinsic Gaussian FWHM $\lesssim 0.5$ 
pixel) or simply very compact (FWHM $>$ 1 pixel).  In cases where a point 
source is not evident in the surface-brightness profile, we determined upper 
limits to the point-source magnitudes that would have resulted in 1$\sigma$ 
detection. We follow the use of constant $\chi^2$ boundaries (Press et al. 
1992) in order to set the upper limits. Adopting the usual convention,
$\chi^2$ is defined as the sum over all pixels of the squares of the 
deviations between data and model image, weighted by the uncertainties.
We first obtained the best-fit model which corresponds to the minimum $\chi^2$
of the fit for the galaxy without a point source. The fitting was then redone by 
introducing point sources of fixed magnitude and FWHM, but allowing all the 
galaxy parameters to vary. This results in convergence back to the best-fit 
model for the galaxy, but with increased $\chi^2$ for the fit. We adjusted the 
point-source magnitude in each trial until the $\chi^2$ increased by an amount 
equal to the reduced $\chi^2$ of the best fit. One of the main concerns when
using this method to obtain upper limits is that the data points are not 
independent, due to the influence of the PSF. We performed a simple test 
using PSFs of different widths and found that the errors on the estimated 
magnitudes are less than $\lesssim$ 0.2 magnitudes. The upper limits on the 
point-source magnitudes, along with all other fitted parameters, are given in 
Table~2.  As expected, the limits are brighter for power-law galaxies compared 
to core galaxies.

The presence of nuclear dust lanes can lower the surface brightness over a 
small range of radii close to the nucleus, often mimicking the presence of a
point source in the surface brightness profile (eg; NGC 4150, 4261, 4374, 5273,
%
%%%%%%%%%%%%%%%%%%%%%%%%%%%%%%%%%%%%%%%%%%%%%%%%%%%%%%%%%%%%%%%%%%%%%%%%%%%
%\vskip 0.3cm
%
%%good bounding box = default
%\figurenum{2}
%\psfig{file=fig2.ps,width=8.5cm,angle=0}
%\figcaption[fig2.ps]{
%Comparison of aperture photometry derived from NICMOS and ground-based
%images.  The measurements have been made with $\sim$10\asec~diameter
%apertures.  The solid line denotes equality.  The average difference
%between the two sets of data is $\langle{\Delta m_H}\rangle\,=\,m_H(HST)\,-\,
%m_H({\rm ground})$ = $-$0.006$\pm$0.11 mag.
%\label{fig2}}
%\vskip 0.3cm
%%%%%%%%%%%%%%%%%%%%%%%%%%%%%%%%%%%%%%%%%%%%%%%%%%%%%%%%%%%%%%%%%%%%%%%%%%%%%
%%
%\noindent
and 7743). In such cases, we made best attempts to correct for the effects 
of dust. The fit was first performed on the observed images without 
masking the dust features. The residual images were then used to locate 
the dust features, and masks were generated. The fit was subsequently 
repeated using the dust masks. 
If the best-fit inner slope cannot account for the excess light in the 
center, we include a central point source. The effect of dust on the 
surface-brightness profile is very prominent in the case of NGC 4150,
and careful masking was essential to remove the artificial nucleation 
caused by the strong, nuclear dust lane.

We evaluate the quality of the 2-D fit in two ways.  First, we perform 
isophote fits on the observed image and compare them with the 2-D fits.
This is illustrated in the surface-brightness profiles of Figure~1, where the 
solid points show the profile derived from the observed image, the dotted 
curve corresponds to the best-fitting 2-D Nuker profile for the bulge, and the 
solid curve represents the bulge profile plus an optional additional point 
source.  It is evident that the 2-D modeling by GALFIT reproduces the observed 
azimuthally averaged profile very well, in nearly all cases.  The second, more 
straightforward way to judge the quality of the fit is to look directly at the 
residual image, formed by subtracting the original image from the 2-D model.  
This is shown in the right grayscale panel of each galaxy in Figure~1.  The 
residual image very effectively highlights the shortcoming of a 
single-component model.  It can be readily seen that the Nuker function is 
not an adequate representation of the central profiles in some objects, 
particularly those with strong stellar disks.  For NGC 3593 and NGC 4111, for 
instance, the single Nuker fit is a gross oversimplification given their 
complex morphologies.  In NGC 3593 (Fig.~1{\it f}) the fitting is 
affected by the presence of chaotic star-forming regions and associated 
dust, while the surface-brightness profile of NGC 4111 (Fig.~1{\it g}) has a 
number of very disky and elliptical components.  In situations where the 
central point source is strong or the bulge profile is very peaked, Tiny Tim 
models the PSF cores well, but it may not reproduce all the fine structures of 
the diffraction rings and spikes (see, e.g., NGC 404, Fig.~1{\it a}; NGC 5273, 
Fig.~1{\it m}; NGC 5548, Fig.~1{\it n}).  The residual image further 
accentuates faint, high-spatial frequency features such as dust lanes and 
star clusters.
%
%%%%%%%%%%%%%%%%%%%%%%%%%%%%%%%%%%%%%%%%%%%%%%%%%%%%%%%%%%%%%%%%%%%%%%%%%%%
%\vskip 0.3cm
%
%%good bounding box = default
%\figurenum{2}
%\psfig{file=fig2.ps,width=8.5cm,angle=0}
%\figcaption[fig2.ps]{
%Comparison of aperture photometry derived from NICMOS and ground-based 
%images.  The measurements have been made with $\sim$10\asec~diameter 
%apertures.  The solid line denotes equality.  The average difference 
%between the two sets of data is $\langle{\Delta m_H}\rangle\,=\langle m_H(HST)\,-\,
%m_H({\rm ground})\rangle$ = $-$0.006$\pm$0.11 mag.
%\label{fig2}}
%\vskip 0.3cm
%%%%%%%%%%%%%%%%%%%%%%%%%%%%%%%%%%%%%%%%%%%%%%%%%%%%%%%%%%%%%%%%%%%%%%%%%%%%%
%%

As a test of our method, we applied GALFIT to the F555W WFPC2 image of NGC 221 
(M32) and the F547M image of NGC 3379, both of which have published 1-D Nuker 
fits.  We performed the Nuker fit using our 2-D modeling procedure for the 
region within 1$^{\prime\prime}$ of the galaxy center for M32, similar to the 
region used by Lauer et al. (1998). The fitted parameters ($\alpha$ = 2.13, 
$\beta$ = 1.47, $\gamma$ = 0.47, $\mu_{b}$ = 12.93, and $r_{b}$ = 0.47) are in 
good agreement with the published 1-D results (see Table~3). We also performed 
the fit over a larger radius ($r$ \lax 10\asec), similar to the region used for 
our NICMOS images.  The results ($\alpha$ = 4.13, $\beta$ = 1.24, $\gamma$ = 
0.51, $\mu_{b}$ = 12.70, and $r_{b}$ = 0.38) closely agree with the parameters
derived using NICMOS images (Table 2). In the case of NGC 3379, the Nuker
fit for the region within the central 10$^{\prime\prime}$ radius on the
optical image yields $\alpha$ = 1.74, $\beta$ = 1.44, $\gamma$ = 0.21,
$\mu_{b}$ = 16.11, and $r_{b}$ = 1.86, which again agree with the published 
fit parameters. From these and other similar tests, we conclude that the fit 
parameters do vary depending on the region chosen for the fit, but that for 
any given region, the GALFIT results closely match the parameters derived 
using 1-D fits.  

\subsection{Estimation of Photometric Accuracy}

We performed aperture photometry in order to compare NICMOS magnitudes
with ground-based NIR measurements compiled by de~Vaucouleurs \& Longo (1988). 
In most cases the ground-based photometry was done with relatively large 
apertures which are beyond the area covered by the NICMOS images, making a 
direct comparison of aperture magnitudes difficult. But for 13 galaxies,
comparison is possible (aperture diameter $\approx 10^{\prime\prime}$), 
and the results are shown in Figure~2.  The average difference between the two 
sets of data is 
$\langle{\Delta m_H}\rangle\,=\langle m_H(HST)\,-\,m_H({\rm ground})\rangle$ =
 $-0.006\pm0.11$ mag.  This level of agreement is consistent with the 
assessment of Stephens et al. (2000), who found 
$\langle{\Delta m_H}\rangle\,=\,0.048\pm0.16$ mag.

%\vskip 1.0cm

\section{Morphology of the Inner Regions}          %%4

\subsection{Dust Features}

Dust in early-type galaxies provides vital clues to understanding their 
evolutionary history. Although elliptical galaxies were once believed to be 
dust-free systems on the basis of the Hubble classification scheme, it was 
realized from ground-based studies that some of these systems show evidence 
for dust (e.g., Hawarden et al. 1981; Sadler \& Gerhard 1985).  The material 
revealed in the dust lanes may be internally generated through stellar mass 
loss, or its origin may be external, such as accretion from interactions 
(Forbes 1991; Knapp, Gunn, \& Wynn-Williams 1992).  Nuclear dust features 
found within the central few hundred parsecs are particularly interesting from 
the standpoint of their dynamical state and their possible relation with 
central activity.  Because of the short dynamical timescales in the center 
(\lax $10^{8}$ yr), the dust should settle in a plane of the galaxy in which 
stable orbits are allowed. On the other hand, if the dust is not settled, it 
is likely to have been acquired recently from an external source. Nuclear dust 
lanes are routinely detected in {\it HST}\ optical images of early-type 
galaxies (Lauer et al. 1995; Forbes et al. 1995; van~Dokkum \& Franx 1995; 
Verdoes Kleijn et al. 1999; de~Koff et al. 2000; Tomita et al. 
%
%%%%%%%%%%%%%%%%%%%%%%%%%%%%%%%%%%%%%%%%%%%%%%%%%%%%%%%%%%%%%%%%%%%%%%%%%%
%\vskip 0.3cm
%
%good bounding box =  100 5 450 750
%\begin{figure*}[t]
%\figurenum{3}
%\centerline{\psfig{file=fig3_rev.ps,width=18.0cm,angle=270}}
%\figcaption[fig3.ps]{
%Dependence of the inner cusp slope $\gamma$  on ({\it a}) absolute
%magnitude of the galaxy ($M^0_{B_T}$) and ({\it b}) break radius ($r_b$).
%The distribution of $\gamma$ is shown in panel ({\it c}).  {\it Filled}\
%symbols represent core galaxies, and {\it open }\ symbols denote power-law
%galaxies.  The galaxies from this study are plotted as {\it circles},
%those from Faber et al. (1997) are shown as {\it triangles}, and those from
%Rest et al. (2001) as {\it squares}.  Objects in our sample which fall in the
%``gap region'' (0.3 $< \gamma <$ 0.5) appear as {\it asterisks}, while those
%from Rest et al. (2001) are shown by {\it plus} signs. Two galaxies which
%fall in this region from Faber et al. (1997) are shown by {\it crosses}.
%\label{fig3}}
%\end{figure*}
%\vskip 0.3cm
%%%%%%%%%%%%%%%%%%%%%%%%%%%%%%%%%%%%%%%%%%%%%%%%%%%%%%%%%%%%%%%%%%%%%%%%%%%
%
%\noindent
2000; Tran et al. 2001). The $V$-band WFPC images of early-type
galaxies analyzed by 
Lauer et al. (1995) revealed dust features in 28\% of the sample, while 
43\% of galaxies in the study by Rest et al. (2001) showed dust disks and
filaments. Interestingly, Rest et al. (2001) find that nuclear dust disks, 
when present, reside mostly in power-law galaxies.  The frequency of 
occurrence of dust lanes appears to be connected with AGN activity. Van~Dokkum 
\& Franx  (1995) and Verdoes Kleijn et al. (1999) find that at least 75\% of 
radio-loud elliptical galaxies contain dust features in their $V$-band 
images.  Analysis of WFPC2 images of 3CR radio galaxies revealed dust features 
in one-third of the sample, with a correlation between the morphological 
distribution of dust and the Fanaroff \& Riley (FR; 1974) classification 
(de~Koff et al. 2000). When present, dust in FR~I sources is distributed in 
small-scale ($\sim$2.5 kpc), sharply defined disks whose major axes often lie 
perpendicular to the radio jets.  FR~II galaxies, by contrast, display dust 
structures with a variety of sizes and morphologies.

Early-type galaxies appear largely smooth in the NIR.  Only $\sim$25\% of the 
galaxies in our sample show evidence for dust in the $H$ band. Some galaxies 
known to be dusty in \hst\ optical images (e.g., NGC 4278, 4589, and 
7626) display smooth morphologies in the NICMOS images, illustrating that the 
NIR images are relatively dust free and can provide more accurate 
surface-brightness profiles. For instance, the profile of NGC 524 in the 
optical is significantly affected by dust (Kormendy 1985a; Lauer et al. 1995);
its dust lanes form concentric shells and are clearly visible in the optical 
images of Lauer et al. (1995).  By contrast, they are barely visible in the 
residual image in Figure~1{\it b}. On the other hand, NGC 5838 has a very 
prominent dust lane that encircles the nucleus and causes severe obscuration 
of the inner regions even in the NIR. The WFPC2 images of this galaxy reveal 
another concentric dust ring located exterior to that seen in Figure~1{\it n}.
Nuclear dust features are also seen in the case of NGC 1052, 4150, and 4374, 
all of which host emission-line nuclei at the center. NGC 3593 shows 
considerable amounts of dust and star formation activity and exhibits 
a chaotic morphology in its center.

Since the NIR images are less sensitive to dust, exploring the relationship 
between dust and AGN activity in our sample would have to be based on the 
optical images. Dust properties for most of our sample galaxies have been 
discussed in previous works (van~Dokkum \& Franx 1995; Tomita et al. 2000; 
Quillen, Bower, \& Stritzinger 2000). For those galaxies that do not have 
published color maps or residual images, we retrieved the optical images from 
the {\it HST}\ archive. Only two galaxies in our sample, NGC 4026 and NGC 
4417, have not been observed using {\it HST}\ in the optical. Based on 
examination of the archival optical images and other published results, we 
find that dust occurs in 60\% of our sample galaxies. Almost all galaxies with 
nuclear activity have dust in the central regions, the only exceptions being 
NGC 474 and NGC 5982. None of the nine galaxies with pure absorption-line 
nuclei show evidence for dust. This reinforces the results from previous 
studies on the association between dust and AGN activity.  It also strengthens 
another related, emerging trend: dust absorption is invariably coupled with
optical line emission in nuclear regions (Verdoes Kleijn et al. 1999; Pogge 
et al. 2000).
 
\subsection{Nuclear Stellar Disks}

Strong nuclear stellar disks are distinctly seen in the residual images of 
NGC 2685, 3115, 4026, 4111, and 4417. NGC 3384 and NGC 3900 have weak disks, 
as seen on the residual images and as reflected in the positive $B_4$ values. 
The presence of disks in these galaxies is not surprising given their S0 
Hubble type.  Two ellipticals --- M32 and NGC 821 --- show mild disklike 
characteristics as well.  As was noted by Michard \& Nieto (1991) from their 
ground-based images, M32 tends to have positive $B_4$ values within 
$r\,\approx$ 5\asec; they further remark that the inner disky isophotes become 
more evident at $\sim$1 \micron.  However, the
%
%%%%%%%%%%%%%%%%%%%%%%%%%%%%%%%%%%%%%%%%%%%%%%%%%%%%%%%%%%%%%%%%%%%%%%%%%%
%
%%good bounding box = 40 260 590 540
%\begin{figure*}[t]
%\centerline{\psfig{file=table3_v4.ps,width=17.5cm,angle=0}}
%\end{figure*}
%
%%%%%%%%%%%%%%%%%%%%%%%%%%%%%%%%%%%%%%%%%%%%%%%%%%%%%%%%%%%%%%%%%%%%%%%%%%
%
%\noindent
WFPC2 $V$ and $I$ images of Lauer et al. (1998)
do not exhibit significant $B_4$ values or any presence of 
an inner disk of red stars.  The residual image of NGC 821 reveals a weak 
disklike feature, and the $B_4$ parameter has small positive values over most 
of the semi-major axis length (see also Lauer 1985b). Both M32 and NGC 821 
possess steeply rising surface-brightness profiles similar to S0 galaxies. The 
frequency of stellar disks in our sample (21\%) is higher than that found by 
Lauer et al. (1995; 13\%).  Rest et al. (2001) find embedded stellar disks 
in a significantly larger fraction of their sample (51\%), which is 
dominated by power-law galaxies.

\vskip 0.3cm

\section{Surface-Brightness Profile Types}    %%5

\subsection{Inner Slopes}

Following the criteria proposed by Lauer et al. (1995), we classify a galaxy 
as core type when there is a significant flattening in the slope of the outer 
power-law profile to an inner slope $\gamma$ which is less than 0.3. Power-law 
galaxies are defined by $\gamma\,>$ 0.5.  Our sample has 13 core galaxies and 
17 power-law galaxies. In our sample, power-law profiles are mostly associated 
with S0 galaxies, except for M32 and NGC 821, which are ellipticals. 
Interestingly, both of these galaxies show evidence for disky isophotes 
(\S~4.2). NGC 404 and NGC 524 are the only S0 galaxies with core-type 
profiles; as discussed in \S~7.1, NGC 404 deviates strongly from the core 
fundamental plane, and it is very likely to have been misclassified. In the 
case of NGC 5548, its classification as a core galaxy is somewhat uncertain 
due to the extremely bright Seyfert nucleus, whose dominant diffraction rings 
contaminate the bulge profile even beyond $r\,\approx$ 1$^{\prime\prime}$.
Out of the seven galaxies with detected stellar disks, all have 
power-law surface-brightness profiles, except for NGC 4111 whose
classification is highly uncertain because the bulge cannot be modeled by 
a single component. In order to see how the isophotal shapes relate to the 
profile type, we adopt the criteria $B_4$ $\geq$ 0.01 for disky galaxies
and $B_4$ $\leq$ $-$0.01 for boxy galaxies. All nine galaxies with disky 
isophotes have power-law profiles except NGC 7457, which has an intermediate 
inner slope. Two galaxies with boxy isophotes have core-type surface-brightness 
profiles. Among the 15 galaxies with pure elliptical isophotes, nine are 
core-type, three have intermediate slopes and three are power laws. 
The remaining galaxies show mixed isophotal shapes due to the presence of dust.

Faber et al. (1997) found that there is a clear dichotomy in the profile
types with the luminosities of the galaxies. Core galaxies are luminous 
systems with $M_V$ \lax $-$20.5 mag;  power-law galaxies are fainter and 
extend up to $M_V$ $>$ $-$22.0 mag. Both profile types overlap in the 
magnitude range $-$20.5 \lax $M_V$ \lax $-$22.0. 
Within the range of resolvable break radii 
occupied by core and power-law galaxies, there was a clear absence of galaxies 
with inner-slope values from 0.3 to 0.5. The galaxies 
occupied distinct regions when $\gamma$ was plotted against log $r_{b}$, with 
a clear gap in the above range.  The core galaxies in our sample are luminous 
systems with boxy or pure elliptical isophotes, while power-law galaxies are 
less luminous and have disky isophotes, in agreement with the results of 
Faber et al. (1997). 

The addition of our sample to that of Faber et al. somewhat blurs the sharp
boundary between core and power-law galaxies (Fig.~3).  Although the 
distribution of $\gamma$ still appears bimodal, four galaxies in our sample 
(NGC 474, 5273, 7457, and 7626) occupy the gap region\footnote{Using early 
\hst\ observations obtained with WFPC, Lauer et al. (1991) concluded that the 
central profile of NGC 7457 was consistent with a $\gamma\,\approx\,-1.0$ 
power law with an additional superposed pointlike nucleus. However, Lauer et 
al. were only able to model the strongly aberrated PSF in a preliminary 
fashion, and we do not consider their result to be in serious conflict with 
ours.}. Rest et al. (2001) provide Nuker fits of the major-axis and minor-axis
profiles of 57 early-type galaxies and have shown that the cusp slopes along
the two axes are similar. Including the inner slopes for the major
axis fits from the sample of Rest et al. (2001) further emphasizes the 
continuous distribution of inner slopes.  

Are there any systematic errors in our profile fitting which 
could have caused some of the galaxies to fall in the gap region?  
Three of the galaxies (NGC 474, 5273, and 7457) have relatively strong, 
pointlike nuclei, and it is possible that the 
%
%%%%%%%%%%%%%%%%%%%%%%%%%%%%%%%%%%%%%%%%%%%%%%%%%%%%%%%%%%%%%%%%%%%%%%%%%%
%\vskip 0.3cm
%
%%good bounding box = 150 10 450 750
%\begin{figure*}[t]
%\figurenum{4}
%\centerline{\psfig{file=fig4.ps,width=18.0cm,angle=270}}
%\figcaption[fig4.ps]{
%Comparison of the values of ({\it a}) $\gamma$, ({\it b}) $\beta$, and
%({\it c}) $r_b$ derived from NICMOS images with those derived from optical
%images. The solid line denotes equality.  Filled circles correspond to
%NGC 524, NGC 4589 and NGC 7626, which are known to have dust in the central
%regions as seen on optical images.  The central profiles of these objects
%most likely were depressed in the optical by dust extinction, leading to
%$\gamma$(opt) $\approx$ 0.
%\label{fig4}}
%\end{figure*}
%\vskip 0.3cm
%%%%%%%%%%%%%%%%%%%%%%%%%%%%%%%%%%%%%%%%%%%%%%%%%%%%%%%%%%%%%%%%%%%%%%%%%%%%
%
%\noindent
nuclei may have affected the determination of the inner-profile
slope.   We do not believe this to be the 
case.  Our tests indicate that the typical errors\footnote{The uncertainties 
represent the 68\% confidence intervals, which we estimated from the use of 
constant $\chi^2$ boundaries (see \S~3.2).} on $\gamma$ are only $\pm$0.03.  
The only exception is NGC 5273, for which an error on $\gamma$ of $\pm$0.20 is 
possible.  NGC 7626 itself has no pointlike nucleus, with an upper 
brightness limit of 
$m_H^{\rm nuc}$ \gax 19 mag (Table 2).  Another possibility is that the 
intrinsic profiles of these objects are truly of the power-law variety 
($\gamma\,>\,0.5$) which, because of exceptional amounts of extinction, even 
at 1.6 \micron, have been artificially depressed to shallower values, thereby 
shifting them into the gap region.  Again, except for NGC 5273, there is no 
evidence that this is so.  NGC 7626 does contain a kinematically distinct 
core (it was included in the study of Carollo et al. 1997), but this 
kinematic attribute is irrelevant to the fidelity of our profile fitting.  
Carollo et al. (1997) obtained $\gamma\,\approx\,0$ for NGC 7626 and 
classified it as a core type.  Their flat inner slope, however, was very
likely caused by the dust lane present in the optical image (see also \S~5.2).  
Quillen et al. (2000) analyzed the $H$-band image and derived
$\gamma\,=\,0.46$.  Our analysis of the same data yielded a slightly shallower
slope of $\gamma\,=\,0.36$.

%\vskip 1.0cm

\subsection{Comparison with Previous Studies}

It is of interest to compare our results with published material, since our 
2-D fitting method differs fundamentally from methods employed in previous 
studies, and since the majority of the existing work is based on imaging at 
optical wavelengths.  Table 3 compares the Nuker-law parameters for a subset 
of 13 galaxies from the present work with those obtained from published 
studies based on WFPC and WFPC2 $V$-band images.  The inner-cusp slopes 
$\gamma$ and the break radii $r_b$ are in good agreement for most galaxies 
(Fig.~4).  This indicates that, to first order, there are no gross systematic 
differences between our results and those of others.  It also suggests that 
there are no strong color gradients between $V$ and $H$ interior to the break 
radius.  By contrast, the majority of the outer slopes $\beta$ determined from 
the NIR images tend to be {\it steeper}\ than the optical values.  As the 
outer slopes are largely insensitive to the detailed treatment of the PSF, 
this result must reflect a genuine difference in the profiles between the two 
bands, one that can most plausibly be attributed to a positive $V-H$ gradient 
toward smaller radii.  

A few cases deserve special attention.  NGC 524, 4589, and 7626 (shown as 
filled circles in Fig.~4) appear consistently as outliers, most notably in the 
$\gamma$ and $\beta$ plots.   The discrepancies between the optical and NIR 
can be ascribed readily to the effects of dust, since all three galaxies show 
clear evidence for dust lanes in the optical images (Lauer et al. 1995; 
Carollo et al. 1997).  Extinction causes the inner slope to appear relatively 
flat in the optical, such that $\gamma({\rm opt})\,\approx$ 0 for all three 
cases, but $\gamma({\rm NIR})$ systematically exceeds $\gamma({\rm opt})$, by 
quite a significant margin in NGC 7626.  The prominent stellar disk in the S0 
galaxy NGC 3115 makes the modeling using the Nuker function difficult; its 
break radius is 85\% larger in the optical than in the NIR.  The differences 
in the Nuker parameters for M32 between our work and that of Lauer et al. 
(1998) is mainly due to the choice of the fitting radius, as discussed in 
\S~3.2.

Quillen et al. (2000) investigated the structural parameters of a sample of 
27 elliptical galaxies observed with NICMOS in the $H$ band, as in our 
study.  Their method of analysis differs from ours.  They fitted a Nuker law 
to the 1-D surface-brightness profiles, which were derived from deconvolved 
images.  No effort was made to account for pointlike nuclei.  Twelve of the 
galaxies overlap with our sample, and for these we find significant systematic 
differences in the derived Nuker-law parameters.  The average difference 
(and standard deviation) between our values and those of Quillen et al. 
are as follows: 
$\langle{\Delta \mu_b}\rangle\,=\,0.64\pm0.92$, 
$\langle{\Delta r_b}\rangle\,=\,0.071\pm0.33$, 
$\langle{\Delta \alpha}\rangle\,=\,-0.89\pm2.00$,
$\langle{\Delta \beta}\rangle\,=\,0.14\pm0.18$, and 
$\langle{\Delta \gamma}\rangle\,=\,-0.098\pm0.073$.   
The systematic discrepancies are most simply explained by the treatment of 
nucleated galaxies.

\section{Central Compact Sources and AGNs}    %%6

Unresolved central nuclei are present in 14 out of the 33 sample galaxies.  
Excluding NGC 3593, 4111, and 5548, whose profile types are uncertain,
we find that the fraction of galaxies containing unresolved nuclei is 
roughly equal between core (38\%) and power-law systems (35\%). What is the 
nature of 
%
%%%%%%%%%%%%%%%%%%%%%%%%%%%%%%%%%%%%%%%%%%%%%%%%%%%%%%%%%%%%%%%%%%%%%%%%%%%
%\vskip 0.3cm
%
%%good bounding box = 20 120 610 700
%\begin{figure*}[t]
%\figurenum{5}
%\centerline{\psfig{file=fig5_rev.ps,width=18.0cm,angle=0}}
%\figcaption[fig5.ps]{
%Central-parameter relations for core galaxies. The core parameters ($r_{b}$
%and  $\mu_{b}$) are well correlated with the central stellar velocity
%dispersion ($\sigma_0$) and, with greater scatter, with the total $B$-band
%absolute magnitude ($M^0_{B_T}$).  Panels ({\it a}) and ({\it b}) plot only
%data from this study ({\it circles}), whereas panels ({\it c}) and ({\it d})
%include measurements from Faber et al. (1997; {\it triangles}) and
%Rest et al. (2001; {\it squares} )
%The outliers NGC 404 and NGC 4636 are labeled.
%\label{fig5}}
%\end{figure*}
%\vskip 0.3cm
%%%%%%%%%%%%%%%%%%%%%%%%%%%%%%%%%%%%%%%%%%%%%%%%%%%%%%%%%%%%%%%%%%%%%%%%%%%
%
%\noindent
these unresolved sources?  Are they AGNs or compact nuclear
star clusters?  How does their occurrence relate to the large-scale 
properties of the host galaxies?  The current statistics are too meager 
to address these issues 
meaningfully, and we merely point out a few noteworthy trends. Eleven of the 
14 nucleated sources are spectroscopically classified as AGNs or closely 
related objects (eight LINERs and three Seyferts), and 
nearly all (5/6) of the ``type~1'' nuclei (those with visible broad emission 
lines) show evidence for point sources. Of the remaining objects, two have 
a pure absorption-line spectrum, and one is an \hii\ nucleus.  The majority 
of the nuclei, therefore, appear to be associated with active galaxies, and 
it is reasonable to postulate that the 1.6 \micron\ emission may be nonstellar
in origin.  Lauer et al. (1995) find compact nuclei in 35\% of their sample;
they occur preferentially in power-law galaxies and are not associated with
nonstellar activity. The only two cases where the central nucleus is an AGN 
turn out to be core-type galaxies. Rest et al. (2001) adopt more conservative
criteria for identifying nuclei in their sample to avoid false detections
arising from artifacts caused by deconvolution and the effects of dust.  They 
find a much lower detection rate of 13\%; only two out of the nine nucleated 
galaxies have core-type profiles, both of which show evidence for AGN activity.

Our statistics on the incidence of compact nuclei are affected
%
%%%%%%%%%%%%%%%%%%%%%%%%%%%%%%%%%%%%%%%%%%%%%%%%%%%%%%%%%%%%%%%%%%%%%%%%%%%
%\vskip 0.3cm
%
%%good bounding box = 15 140 540 680
%\figurenum{6}
%\psfig{file=fig6.ps,width=8.5cm,angle=0}
%\figcaption[fig6.ps]{
%Fundamental-plane relation between $\mu_{b}$ and $r_{b}$ for core galaxies.
%The outliers NGC 404 and NGC 4636 are labeled.
%\label{fig6}}
%\vskip 0.3cm
%%%%%%%%%%%%%%%%%%%%%%%%%%%%%%%%%%%%%%%%%%%%%%%%%%%%%%%%%%%%%%%%%%%%%%%%%%%%%
%\noindent
in part by resolution effects and should be considered lower
limits. A handful of our
galaxies were observed with the NIC3 camera, whose pixel scale is 2.5 times
coarser than that of NIC2.  The ability to discern very faint, photometrically
distinct nuclei depends critically on image resolution.  This is most 
dramatically illustrated in the case of NGC 4278, for which data are available 
from both cameras.  The NIC3 image shows no clear indication of a central 
point source, but in the higher resolution NIC2 image the compact nucleus 
emerges unambiguously from the underlying galaxy profile.  Similarly, the NIC3 
image of NGC 3115 used in the present study shows no evidence for the central 
star cluster found in WFPC2 images by Kormendy et al. (1996).

It is of interest to note an apparent dependence of the incidence of nuclear 
ionized gas on profile type.  Nearly all of the core-type galaxies in our 
sample (12/15 or 80\%) exhibit detectable optical line emission within the 
central 2\asec$\times$4\asec\ (few hundred pc) above a spectroscopic 
equivalent-width limit of $\sim$0.25~\AA\ (Ho et al. 1997a).  In accordance
with known trends of spectral class with Hubble type and galaxy luminosity 
(Ho et al. 1997b), the nuclei are classified as AGNs or closely related 
objects, the majority of the LINER variety.  Of the 17 galaxies with 
power-law profiles (including the few in the gap region), only 11 (65\%) 
have detectable nebular emission, all of which are also classified as AGNs. 
The marginally higher detection frequency of nuclear line emission among 
core-type systems comes somewhat as a surprise, given that the central 
regions of ellipticals in general emit weaker optical line emission and 
contain a smaller mass of ionized ($\sim 10^4$ K) gas than do lenticulars 
(Ho 1999b).  The apparent conflict may simply reflect small-number statistics.

\section{Discussion}         %%7

\subsection{Fundamental-Plane Relations for Galaxy Cores}

Elliptical galaxies and bulges obey global correlations between size ($r_e$), 
surface brightness ($\mu_e$), velocity dispersion ($\sigma_e$), and luminosity 
--- the ``fundamental plane'' (Faber et al. 1987; Dressler et al.  1987; 
Djorgovski \& Davis 1987; Bender, Burstein, \& Faber 1992).  Luminous
galaxies have larger effective radii (lower central densities), lower surface
brightnesses, and higher velocity dispersions (Kormendy 1985b). These scaling
relations have important implications for theories of galaxy formation, which
must successfully reproduce them.  Faber et al. (1997) have discussed the
existence of a fundamental plane in the ($r_{b}$, $\mu_{b}$, $\sigma_0$)-space 
for galaxy cores, analogous to the one known in the ($r_{e}$, $\mu_{e}$, 
$\sigma_{e}$)-space on larger scales. In Figures 5 and 6, we reproduce the 
fundamental-plane relations for the core galaxies in our sample.  The break 
radii obtained for power-law galaxies are not physically meaningful, and hence 
are not considered here, although these galaxies may have genuine cores on 
scales smaller than we can resolve in these observations.  The 
log~$r_b$--$M^0_{B_T}$ (Fig.~5{\it c}) and log~$r_b$--$\sigma_0$ 
(Fig.~5{\it d}) plots incorporate additional, nonoverlapping core galaxies 
from the sample of Faber et al. (1997) and Rest et al. (2001)\footnote{ For 
consistency with our sample, the absolute 
magnitudes plotted in Figures~3 and 5 were derived from $B^0_T$ magnitudes given
in de~Vaucouleurs et al. (1991) using distances from Tully (1988) when available, 
and otherwise from heliocentric radial velocities listed in the 
NASA/IPAC Extragalactic Database (NED), using $H_0$ = 75 \kms\ Mpc$^{-1}$.}. 
As demonstrated in \S~5.2, our NICMOS-based measurements of $r_b$ agree well 
with values derived from WFPC and WFPC2 images.  

As Faber et al. (1997) have found in the optical, our study shows that the 
core parameters of elliptical and S0 galaxies similarly follow the 
fundamental-plane relations in the NIR, again by direct analogy to well 
established scaling relations on global scales (e.g., Pahre, de~Carvalho, \& 
Djorgovski 1998; Scodeggio et al. 1998; Mobasher et al. 1999).  The 
surface brightness at the break radius and the break radius correlate well 
with the central velocity dispersion (Fig.~5{\it b}, 5{\it d}), but having 
greater scatter with total galaxy luminosity (Fig.~5{\it a}, 5{\it c}) --- 
more luminous ellipticals with higher velocity dispersions have larger cores 
and lower surface brightnesses.  A tight correlation exists between $\mu_{b}$ 
and $r_{b}$ (Fig.~6), similar to the $\mu_{e}$--$r_{e}$ relation defined 
at the effective radius. Excluding the two outliers, a least-squares fit 
gives $\mu_{b}\,\propto\,r_{b}^{1.80\pm0.15}$, with an rms scatter of 0.24 
about this fit. 

Two galaxies --- NGC 404 and NGC 4636 --- deviate from the norm by their 
unusually low surface brightness.  NGC 404 is a nearby ($D$ = 2.4 Mpc), 
low-luminosity ($M^0_{B_T}\,\approx\,-16$ mag) S0 galaxy which contains a very 
prominent nucleus ($m^{\rm nuc}_H$ = 13.5 mag; $M^{\rm nuc}_H\,=\,-13.4$ mag).  
The nucleus appears as a bright ultraviolet source with signatures of 
young, massive stars (Maoz et al. 1998).  With $\gamma$ = 0.28 and a small 
break radius ($r_{b}$ = 0\farcs40), however, the status of NGC 404 as a core 
galaxy is rather shaky. The inner slope depends critically on the 
decomposition of the bright nucleus, and equally acceptable fits can be 
admitted with $\gamma\,\approx$ 0.28--0.50.  The classification of NGC 4636 as 
a core galaxy, on the other hand, is quite secure ($\gamma$ = 0.13; $r_{b}$ = 
3\farcs44).  The central light distribution is not affected by a nuclear point 
source ($m_H^{\rm nuc}\,>$ 23 mag), and no sign of severe absorption is 
evident in the residual image (Fig.~1{\it m}).  The fitted Nuker parameters, 
moreover, show excellent agreement with those obtained by Faber et al. (1997) 
based on WFPC2 data (Table~3), further suggesting that dust extinction has a 
minimal impact on the photometric parameters.  Thus, the displacement of 
NGC 4636 from the core scaling relations appears to be genuine.

\subsection{Distribution of $\gamma$: Bimodal or Continuous?}

Faber et al. (1997) emphasized the apparent gap in the distribution of the 
inner slopes, with no galaxies having $0.3 < \gamma < 0.5$.   The
distribution of $\gamma$ appeared bimodal. Gebhardt et al. (1996) deprojected 
the surface-brightness profiles of Lauer et al. (1995) to examine the 
luminosity-density profiles, and they showed that the inner slopes of the 
latter also exhibited a bimodal distribution. However, the recent analysis 
of Rest et al. (2001) reveals that almost 10\% of their galaxies have 
asymptotic inner slopes in the range $0.3 < \gamma < 0.5$.  Thus, whether the 
distribution of $\gamma$ is bimodal or not remains somewhat controversial.

In the present study, four galaxies have inner slopes in the ``gap region'' 
(Fig.~3); these include the S0 galaxies NGC 474, 5273, and 7457, and the 
elliptical NGC 7626. The presence of a bright nucleus or dust lanes can affect 
the determination of the inner slope. All three S0 galaxies show evidence for 
bright nuclei.  But as mentioned in \S~5.1, our tests indicate that the 
only case where the point source may be problematic is NGC 5273, whose inner 
slope has an unusually large uncertainty ($\gamma$ = 0.37$\pm$0.20).  The 
1$\sigma$ errors on $\gamma$ are less than 0.03 for the remaining three 
galaxies.  Similarly, apart from NGC 5273, none of the other three galaxies 
show significant dust in the $H$ band that could have affected the 
determination of $\gamma$. 

Faber et al. (1997) and Rest et al. (2001) have discussed the ambiguity 
introduced in the classification of profile types due to distance effects.
An intrinsic core galaxy placed at a large distance is likely to be classified 
as a power-law galaxy if the break radius is close to the resolution limit. 
Similarly, for small values of $\alpha$, the slope changes gradually with 
radius, and the asymptotic value of $\gamma$ may be reached only at radii 
much smaller than the resolution limit. Rest et al. (2001) introduced a new 
parameter, $\gamma^{\prime}$, which is the gradient at 0\farcs1 derived from 
the best Nuker fit, and their classification of the surface-brightness 
profiles is based on this parameter.  Faber et al. (1997) illustrated using 
M31 how distance effects can produce misleading $\gamma$ values.  M31 is 
classified formally as a core galaxy, but it would be classified as a 
power-law galaxy if it were at the distance of the Virgo cluster. Intermediate 
values of $\gamma$, therefore, could result from measuring the inner slope too 
close to the break radius, especially when $\alpha$ is small. In the case of 
the four galaxies in our sample with intermediate $\gamma$, the sharpness of 
the transition from the inner to outer slopes as indicated by $\alpha$ is 
significant, and their break radii are, for the most part, well resolved 
($r_b\,\approx$ 0\farcs33--1\farcs47).  Thus, it does not appear that the 
intermediate inner slopes of these galaxies can be attributed readily to 
distance effects.

Although the introduction of the four intermediate objects is insufficient to 
erase the bimodality of the $\gamma$ distribution, it is of interest to ask 
why previous studies have failed to find such objects. The studies by Lauer et 
al. (1995) and Faber et al. (1997) explicitly avoided galaxies known to be 
dusty. The selection criteria of our study, on the other hand, are more 
general, and do not bias against objects with dust.  Since it is reasonable to 
suppose that cold gas content scales with dust content, one wonders whether 
our less restrictive selection criteria may have uncovered certain galaxies in 
unusual evolutionary states. Although Rest et al. (2001) adopt a different 
definition for the inner slope, it is intriguing that they too find a handful 
of galaxies in the gap region.    

NGC 7626 contains a kinematically distinct core, a possible relic of a former
galaxy interaction or merger event.  According to Carollo et al. (1997), 
however, this class of ellipticals does not exhibit overtly different 
photometric properties on \hst\ scales compared to kinematically normal 
ellipticals.  Carollo et al. also questioned the reality of the supposed 
dichotomy between core and power-law galaxies; they suggested that the trends 
between inner cusp slope and global galaxy properties may instead be 
continuous.  Their arguments, however, were based on the usage of the average 
logarithmic slopes of the nuclear profiles between 10 and 50 pc 
($\langle\gamma_{\rm phys}\rangle$) instead of on the asymptotic inner slope 
of the Nuker function ($\gamma$), with which the dichotomy between core and 
power-law galaxies was originally proposed, and is the one adopted in this 
study.

The remaining three sources with intermediate $\gamma$ are S0 galaxies.  NGC 
5273 is uncertain, as described above.  However, apart from being nucleated, 
itself not an uncommon attribute, neither NGC 474 nor NGC 7457 show any 
particularly noteworthy characteristics.  They are not exceptionally dusty; 
their global optical colors appear typical of S0 galaxies; and their 
central stellar velocity dispersions roughly conform to the Faber-Jackson 
(1976) relation for S0s, indicating normal mass-to-light ratios.  

The present statistics on galaxies with intermediate cusp slopes are clearly
too meager to warrant excessive speculation on their physical nature.  Larger 
samples are needed to place these objects in the context of current formation 
scenarios for power-law and core galaxies (Faber et al. 1997).

\subsection{Isothermal Cores}

A remarkable finding that emerged from high-resolution ground-based studies 
is the virtual absence of isothermal cores in ``normal'' elliptical galaxies 
(Lauer 1985a; Kormendy 1985a, 1985b).  The handful of objects with candidate 
isothermal cores all turn out to be bright cluster galaxies. {\it HST}\ 
studies confirmed that King models with constant-density cores provide a poor 
description of the observed central surface-brightness profiles of 
ellipticals. Galaxies which showed distinct cores had inner profiles with a 
central rising cusp. Lauer et al. (1995) emphasised that even galaxies with 
very shallow cusps in the inner region of surface-brightness profile have 
steeply rising central luminosity-density profiles, thereby ruling out the 
possibility of isothermal cores. 

Six galaxies in our sample formally have $\gamma\, <\, 0.05$; the errors on 
$\gamma$ are small, typically less than $\pm$0.02.  Both NGC 4261 and NGC 4278 
have unresolved nuclei, making their inner slopes less reliable.  
NGC 524 and NGC 4472 have weak central cusps, with $\gamma = 0.03$ and 0.04, 
respectively. (Ferrarese et al. 1994 fitted the optical \hst\ profile of 
NGC 4472 with an isothermal model, but the model does not fit the observed 
profile well, and dust is clearly present in their image.)  Excluding these 
cases, the two remaining objects, NGC 4291 and NGC 4406, are excellent 
candidates for galaxies with isothermal cores.  The core of NGC 4406 is very 
well resolved ($r_{b}$ = 1\farcs00), and the inner slope is identically zero 
(1$\sigma$ error $\pm 0.003$).  The image appears very smooth, showing no 
evidence of a distinct nucleus, strong dust absorption, or other 
irregularities.  Kormendy (1985a) already demonstrated that the central region 
of NGC 4406 is well fitted by an isothermal profile; beyond $\sim$6\asec\ the 
isothermal model is inadequate.  Our analysis, based on observations at a 
redder bandpass and much higher resolution, lends greater confidence to 
Kormendy's finding.  The core in NGC 4291 is slightly more compact ($r_{b}$ = 
0\farcs48), but there is also no reason to suspect that the surface-brightness 
profile might be anomalous, and it has $\gamma$ = 0.02$\pm$0.02.  The case for 
an isothermal core in NGC 4291 
%
%%%%%%%%%%%%%%%%%%%%%%%%%%%%%%%%%%%%%%%%%%%%%%%%%%%%%%%%%%%%%%%%%%%%%%%%%%%
%\vskip 0.3cm
%
%%good bounding box = 150 70 450 710
%\figurenum{7}
%\psfig{file=fig7_rev.ps,width=8.5cm,angle=0}
%\figcaption[fig7.ps]{
%Correlation of nuclear point-source magnitude with ({\it a})
%extinction-corrected H$\alpha$ luminosity and ({\it b}) central stellar
%velocity dispersion.  {\it Filled}\ symbols denote nuclei in core galaxies,
%{\it open} symbols represent nuclei in power-law galaxies, and
%{\it asterisks}\ mark objects with intermediate inner slopes
%(0.3 $< \gamma <$ 0.5).  Upper limits are indicated with arrows. Two objects
%with uncertain nuclear profiles appear as {\it triangles}.
%\label{fig7}}
%\vskip 0.3cm
%%%%%%%%%%%%%%%%%%%%%%%%%%%%%%%%%%%%%%%%%%%%%%%%%%%%%%%%%%%%%%%%%%%%%%%%%%%%
%
%\noindent
is thus somewhat less convincing than in NGC 4406, but we consider it a 
possible candidate nonetheless.  Within the errors, the logarithmic slope 
of the luminosity-density profile is nearly zero in both these galaxies.  
Neither object is a bright cluster galaxy.

A few galaxies in the Faber et al. (1997) sample have flat, central 
surface-brightness profiles with $0 < \gamma < 0.02$ (Fig. 3). 
Only two of them have cores with constant luminosity-density, namely
NGC 1600 and NGC 4889 (Gebhardt et al. 1996).
NGC 4889 is one of the brightest cluster galaxies in Coma, but NGC 1600 is not 
a bright cluster galaxy.  Thus, while isothermal cores are generally thought to 
occur exclusively in bright cluster galaxies, a minority of objects appear to 
violate this trend.

\subsection{Properties of Central Unresolved Sources}

Until {\it HST}\ images became available it was difficult to identify and 
quantify photometrically distinct nuclei embedded in the bulges of nearby 
galaxies.  The difficulty of obtaining bulge-free magnitudes for AGNs is 
clearly borne out by the lack of reliable luminosity functions for nearby AGNs 
(Krolik 1998).  Two extreme examples in our sample illustrate the point. The 
luminous elliptical NGC 4278 contains a weak ($m_H^{\rm nuc}$ = 17.2 mag), 
intrinsically faint ($M_H^{\rm nuc}$ = $-$12.8 mag) nucleus which is 
overwhelmingly swamped by host-galaxy light.  By contrast, the well known 
Seyfert nucleus in NGC 5548 ($m_H^{\rm nuc}$ = 11.9 mag; $M_H^{\rm nuc}$ = 
$-$22.2 mag) accounts for a significant fraction of the total luminosity.  We 
have applied our 2-D decomposition method self consistently to measure or 
constrain the strength of the nuclear component.  Excluding NGC 5548, detected 
nuclei have magnitudes in the range $m^{\rm nuc}_H$ = 12.8 to 17.4 mag, or 
$M_H^{\rm nuc}$ = $-$12.8 to $-$18.4 mag.  Undetected sources have limits as 
low as $m_H^{\rm nuc}\,\approx$ 23 mag and $M_H^{\rm nuc}\,\approx$ 
$-$6.5 mag.  

The physical nature of the compact nuclei is not well constrained.  All of the 
sources appear unresolved at the resolution of the NICMOS images, which 
corresponds to FWHM diameters of 14 and 18 pc for NIC2 and NIC3, respectively, 
for the sample median distance of 17 Mpc.  These scales are insufficient to 
distinguish between a truly pointlike nucleus and a compact star cluster.  
With a few exceptions, we also have no information on the spectral properties 
of the sources on nuclear scales.  However, as discussed in \S~6, the 
association of the nuclei with AGN classifications (as determined from 
ground-based spectra) suggests that they may indeed be nonstellar in origin.
If this is the case, the NIR continuum emission might scale with the optical
line luminosity (Fig.~7{\it a}).  The two variables are marginally correlated
at the 94\% confidence level according to the generalized Kendall's $\tau$
test (Isobe, Feigelson, \& Nelson 1986).  However, the significance disappears
altogether when we properly account for the mutual dependence of the two
variables on distance; the partial Kendall's $\tau$ test (Akritas \& Siebert
1996) cannot reject with 52\% probability the null hypothesis that there is
no correlation.  The lack of a clear relation between the NIR continuum and
optical line emission, on the other hand, is hardly surprising, since the 
1.6-\micron\ window is far removed from the ionizing part of the spectrum.  
Moreover, the spectral energy distributions of low-luminosity AGNs exhibit a 
diversity of forms (Ho 1999c), rendering any correlation between the NIR and 
far-ultraviolet bands nontrivial.

Finally, we note that the NIR luminosity of the nuclear sources does not 
depend on the central black-hole mass.  Recent observations suggest that 
massive black holes may be ubiquitous in bulge-dominated galaxies (Magorrian 
et al. 1998; Richstone et al. 1998; Ho 1999a) and that their masses are 
tightly correlated with the stellar velocity dispersion of the bulge 
(Ferrarese \& Merritt 2000; Gebhardt et al. 2000).  A plot of $M_H^{\rm nuc}$ 
versus central velocity dispersion reveals no correlation (Fig.~7{\it b}). To 
the extent that the NIR continuum traces emission from the accretion flow, 
this is indicative of the enormous range spanned by the radiative output of 
massive black holes in nearby early-type galaxies.  

\section{Summary}         %%8

We analyzed $H$-band NICMOS images of the central regions of 33 nearby 
early-type (E, S0, and S0/a) galaxies using a new 2-D fitting technique. 
Our method 
models the galaxy bulge or spheroid using the Nuker function and simultaneously 
accounts for the PSF and a possible nuclear point source.  Our main results 
are as follows:

(1) Early-type galaxies have relatively smooth light distributions in 
the $H$ band, showing dust features only in very few cases, mostly associated 
with galaxies hosting emission-line nuclei. The optical $V$-band WFPC2 images 
show evidence for dust absorption in all the galaxies in the sample that have 
nuclear activity, while dust is absent in galaxies with no nuclear line 
emission.

(2) The Nuker parameters derived from the NICMOS data generally 
show good agreement with published values based on optical WFPC2 images.

(3) Consistent with earlier studies, galaxies with distinct cores
tend to be luminous systems with boxy isophotes, whereas less luminous 
galaxies with disky isophotes tend to have power-law surface-brightness 
profiles with no clearly defined break.  However, we do not find a clear
dichotomy in the distribution of the inner cusp slopes of the Nuker function.  
Specifically, there are several objects with $\gamma$ between 0.3 and 0.5, 
a region previously thought to be empty.

(4) Core galaxies obey the core fundamental-plane relations in 
($r_{b}$, $\mu_{b}$, $\sigma_0$)-space, with more luminous, higher velocity 
dispersion objects having larger cores and lower surface brightnesses.  

(5) Two galaxies (NGC 4291 and NGC 4406) have constant central 
surface-brightness profiles.  They are good candidates for having 
constant-density, isothermal cores.  Neither galaxy is in a rich cluster, 
unlike other systems previously found to host candidate isothermal cores.

(6) Approximately half of the galaxies in our sample are nucleated, 
roughly evenly split between core and power-law galaxies.  Photometrically 
distinct nuclei are especially prevalent in ``type~1'' AGNs (those with 
detected broad emission lines). 

(7) The pointlike nuclei have magnitudes in the range $m^{\rm nuc}_H$ = 
12.8 to 17.4 mag, which correspond to $M_H^{\rm nuc}$ = $-$12.8 to $-$18.4 
mag.  We find no significant correlation between the NIR luminosities and 
the optical line luminosities.  The strength of the nuclear NIR emission also 
appears to be unrelated with the central stellar velocity dispersion, an 
indirect indicator of the black-hole mass.

\acknowledgements
This work is funded by NASA LTSA grant NAG 5-3556, and by NASA grants
HST-AR-07527 and HST-AR-08361 from the Space Telescope Science Institute
(operated by AURA, Inc., under NASA contract NAS5-26555). A.V.F.
acknowledges support from a Guggenheim Foundation Fellowship. 
We thank the referee for helpful comments. We are grateful to 
Roeland P.~van~der~Marel for software to treat the NICMOS pedestal effect.  
We made use of the NASA/IPAC Extragalactic Database (NED) which is operated 
by the Jet Propulsion Laboratory, California Institute of Technology, under 
contract with NASA.

%APPENDIX
%\clearpage
\appendix

\section{Notes on Individual Objects}

This Appendix gives a short description of each object. Most of our sample
galaxies have been studied extensively in various contexts.  We do not attempt 
to provide a detailed description of each galaxy, but instead focus on the 
significant morphological features, peculiarities, or any noteworthy
difficulties encountered in the 2-D modeling.

\vskip 1cm

\indent {\it NGC~221 (M32)}. --- 
The significantly positive $B_4$ values within $r\,\approx$ 5\asec\ suggest
that a disky component is present in M32, although the residual image does not 
show clear evidence for it.  As in the optical (Lauer et al. 1998), the galaxy 
is remarkably smooth and featureless in the NIR.  

\indent {\it NGC~404}. ---
Apart from the strong nuclear point source, the NICMOS image of NGC 404 is 
relatively featureless.  The excess positive emission in the center of the 
residual image is due to a slight mismatch between the point source and the 
PSF.  WFPC2 optical images (Pogge et al. 2000) show a spiral pattern extending 
all the way into the nucleus, vague hints of which can be seen in the NIR 
residual image.  Interestingly, the best fit for the 
surface-brightness profile gives $\gamma$ = 0.28, officially in the domain of 
core galaxies.  This is unexpected given the low luminosity and low velocity 
dispersion of the galaxy.  As seen from Figures 5 and 6, NGC 404 deviates
strongly from the locus of core galaxies in the core-parameter relations. It 
is likely that our estimate of $\gamma$ has been affected by the bright nucleus.

\indent {\it NGC~474}. ---
Although NGC 474 is known to be a shell galaxy with obvious signatures of a 
merger remnant at large radii (Turnbull, Bridges, \& Carter 1999), the inner 
regions appear undisturbed, showing no evidence for significant structure or 
dust, in either the $V$ or $H$ bands.  The low-level features in the residual 
image are due to a slightly mismatched PSF for the nucleus and deviations from 
the Nuker-law fit.

\indent {\it NGC~524}. ---
The presence of a concentric dust pattern in NGC 524 causes the inner profile 
to flatten considerably in the optical.  The dust rings are less prominent 
in the NIR, but they can still be seen in the residual map. Our 2-D Nuker fit 
of the $H$-band image gives $\gamma$=0.03$\pm$0.01. Using the same data, 
Quillen et al. (2000) obtained a much steeper slope of $\gamma$ = 0.25.  
We performed the 2-D fit using their derived Nuker parameters and 
find that even though the observed profile is well reproduced for the range 
0\farcs3--5\farcs0, the fit deviates considerably from the observed 
profile in the inner and outer regions. Within $r$ = 0\farcs1, $\gamma$ = 0.25 
makes the inner regions much brighter than observed.

\indent {\it NGC~821}. --- 
No evidence for dust is seen in the archival $V$-band image. 
The features in the 
residual image are due to slight deviations from the Nuker-law fit.  The 
high ellipticity and positive $B_4$ values over most of the semi-major axis 
length suggest that a disk is present.

\indent {\it NGC~1052}. --- 
The faint structure in the residual image is most likely associated with the 
gas and dust complex seen in optical images.  Pogge et al. (2000) argue that 
the morphology of the optical emission is reminiscent of ionization cones 
found in Seyfert galaxies.

\indent {\it NGC~2685}. ---
This S0 galaxy shows a strong disk within $r\approx$ 2\asec.  As evident in 
the residual image, the disk is poorly modeled with a single-component Nuker 
function.  The dust patches in the archival $V$ image are largely absent 
in the NIR, except for the extended filament located $\sim$5\asec\ to the 
northeast of the nucleus.

\indent {\it NGC~3115}. ---
The residual image shows a nuclear disk in addition to the large-scale disk; 
neither is well fitted with the Nuker law. NGC 3115 is known to have a 
$V\,\approx$ 17 mag central star cluster in the optical (Kormendy et al. 
1996).  Our NIC3 image does not detect the cluster, presumably because 
of its low contrast and the poor pixel resolution.  The optical images 
show no dust features (Lauer et al. 1995; Tomita et al. 2000).

\indent {\it NGC~3379}. ---  
The weak dust ring seen in optical images (van~Dokkum \& Franx 1995; Tomita et 
al. 2000) is not present in the NIR.  The features in the residual image are 
due to slight deviations from the Nuker-law fit.

\indent {\it NGC~3384}. ---
The optical images show no dust features (van~Dokkum \& Franx 1995; Tomita et 
al. 2000).  The residual image highlights the weak nuclear and large-scale 
disks, which are poorly fitted by the Nuker law.  The pointlike nucleus is 
also incompletely removed by the PSF.

\indent {\it NGC~3593}. ---
The central region has a very chaotic morphology from the copious dust lanes 
and star-forming regions. The 2-D Nuker fit is obviously corrupted by the dust 
features, and we consider it to be unreliable.  Fortunately, the strong nuclear 
source is seen quite distinctly, and we can measure its magnitude with little 
complication.

\indent {\it NGC~3900}. ---
The archival $V$ image shows patchy dust, but this is not present in the 
NIR image.  As seen from the residual image and the radial variation of the 
$B_4$ parameter, there is a nuclear disk within the inner 0\farcs8. The 
profile of NGC 3900 would probably be better modeled by multiple components. 
To achieve an acceptable fit with the Nuker law, we restricted the fit to 
$r\,\leq$ 3\asec.

\indent {\it NGC~4026}. ---
The strong disk in the center complicates the modeling. There are no obvious 
dust features in the residual image.  No optical image is available.

\indent {\it NGC~4111}. ---
The central region appears to be composed of a point source, a peanut-shaped 
bulge, and a dominant edge-on disk. Since parameterizing this multi-component 
light distribution by a single Nuker law is an extreme oversimplification, 
we use the 2-D fit only as a guide to extract the brightness of the nucleus.  
We consider the specific values of the Nuker parameters to be unreliable.  
The archival 
$V$-band image shows fan-shaped dust features emanating perpendicularly 
from the disk.  This is barely visible in the residual NIR image.

\indent {\it NGC~4143}. ---
The archival $V$-band image shows patchy dust obscuration close to the 
nucleus, qualitatively similar to the low-level features in our
residual image.

\indent {\it NGC~4150}. ---
A strong nuclear dust lane causes considerable obscuration of the central 
region, both in the optical and in the NIR. The dust lane was carefully
masked while doing the 2-D fit. 

\indent {\it NGC~4261}. ---
This galaxy is remarkable for its high degree of boxiness.  The nuclear 
dust disk of NGC 4261 appears very prominently in optical \hst\ images (Jaffe 
et al. 1993; Verdoes Kleijn et al. 1999), but it is much less conspicuous in 
the NIR.  A pointlike nucleus is definitely required to fit the inner 
light profile.  Quillen et al. (2000) obtained a steeper inner cusp
slope because they did not account for the excess emission from the nucleus.

\indent {\it NGC~4278}. ---
The irregular dust patches of the optical images (van~Dokkum \& Franx 1995; 
Carollo et al. 1997; Tomita et al. 2000) are barely visible in the 
NIR residual map.  The central point source is slightly mismatched by the PSF.

\indent {\it NGC~4291}. ---
The residual image does not show any special feature except for a weak 
quadrupole pattern, which results from the boxy outer isophotes (Ebneter, 
Davis, \& Djorgovski 1988). No dust is obviously present in an archival 
$V$-band image.  The surface-brightness distribution steepens with respect to 
the best-fit Nuker profile for $r$ \gax 3\asec; we thus restricted the fit to 
$r\,\leq$ 3\asec.

\indent {\it NGC~4374}. ---
The multiple dust lanes of NGC 4374 (M84) are clearly visible in the residual 
image and have been extensively discussed in the literature (van Dokkum \& 
Franx 1995; Bower et al. 1997; Verdoes Kleijn et al. 1999). Extensive 
H$\alpha$ emission is associated with the dust.  The embedded nucleus, 
detected in the optical (Bower et al. 1997) and in the NIR (this study), 
is significantly reddened by the dust (Bower et al. 1997).

\indent {\it NGC~4406}. ---
This galaxy is well known for its photometric and kinematic peculiarities. 
It exhibits minor-axis rotation and has a kinematically distinct core 
(Forbes et al. 1995). Surface-brightness profiles obtained from ground-based 
observations under good seeing showed that it has an isothermal core 
(Kormendy 1985a).  The \hst\ profile is remarkably flat, both in the optical 
in the NIR; our fit formally gives $\gamma$ = 0.00$\pm$0.003.  Little or 
no color gradient has been measured by Carollo et al. (1997) and 
Tomita et al. (2000), suggesting that the flattening of the central light 
profile is probably intrinsic and not due to dust extinction.  We consider this 
galaxy to be a strong candidate for hosting an isothermal core (\S~7.3). 

\indent {\it NGC~4417}. ---
The 2-D fit is affected by the strong nuclear disk.  No optical 
\hst\ images are available.

\indent {\it NGC~4472}. ---
As in NGC 4406, the surface-brightness profile of NGC 4472 is very flat 
($\gamma$ = 0.04) in the center and the isophotal parameters show large 
variations. Irregular, patchy obscuration affects the center of the optical 
images (van~Dokkum \& Franx 1995; Tomita et al. 2000), but our NIR residual 
image appears quite smooth.  

\indent {\it NGC~4589}. ---
The complex gas and stellar kinematics of NGC 4589 suggest that it is a 
merger remnant (M\"{o}llenhoff \& Bender 1989).  The dust filaments which
traverse the center of the optical image (Tomita et al. 2000; Quillen et al. 
2000) are also visible in our NIR residual map. The light distribution beyond
$\sim$5\asec\ deviates from the Nuker profile, and we confined our fit to
the region interior to this radius.

\indent {\it NGC~4636}. ---
Van~Dokkum \& Franx (1995) find irregular dust lanes in the optical image, but 
our NICMOS image is extremely smooth.  Although the ellipticity and $B_4$ 
parameter are both well behaved, the position angle shows large variations. As 
discussed in \S~7.1, NGC 4636 deviates markedly from the 
fundamental-plane relations for galaxy cores; its surface brightness is too 
low compared to what is expected for galaxies of similar absolute luminosity
and velocity dispersion.  Likewise, its break radius is systematically 
larger than expected, implying an unusually diffuse core.

\indent {\it NGC~5273}. ---
The galaxy hosts a bright Seyfert nucleus.  The dust absorption pattern
seen in the archival optical image corresponds closely to the structures in 
the NIR residual map.

\indent {\it NGC~5548}. ---
The extremely bright Seyfert nucleus dominates over the bulge light of 
NGC 5548 for radii \lax 1\asec\--2\asec.  The Nuker parameters are 
thus quite uncertain.  No useful (unsaturated) optical image was found in the 
archive.  

\indent {\it NGC~5838}. ---
The thick nuclear dust ring partly occults the nucleus and compromises the 
Nuker fit for the bulge.  We were unable to improve the fit by masking out the 
dust ring.  No estimate of the nuclear point source is given.  A second 
concentric dust ring at larger radii is visible on archival optical images, 
and it can be faintly seen in the $H$-band residual image. 

\indent {\it NGC~5982}. ---
There is no evidence for dust or a point source in this galaxy, which is known 
to contain a  kinematically distinct core (Forbes et al. 1995). The isophotal 
structure, however, is very complicated.  The central isophotes are perfectly
circular, but beyond $\sim$1\asec\ they become increasingly boxy.  This 
leads to the quadrupole pattern in the residual image.

\indent {\it NGC~6340}. ---  
The archival $V$-band image shows a faint dust lane cutting across the 
nucleus, but there is no evidence of it in the NIR.  

\indent {\it NGC~7457}. ---
The nuclear compact source is prominent both in the NIR and in the 
optical (Lauer et al. 1991).   There is no evidence for dust in the 
optical images (Tomita et al. 2000).  The faint features in the NIR 
residual image come from a slight mismatch between the model and the 
data in the region $r\,\approx$ 0\farcs5--1\asec.

\indent {\it NGC~7626}. ---
The warped nuclear dust lane seen in the optical images is likely to be the 
culprit for flattening the optical profile toward the center (Carollo et al. 
1997).  The dust lane is absent from our NIR residual image, and we obtain 
a much steeper central cusp slope ($\gamma$ = 0.36 instead of $\gamma$ = 0.0).

\indent {\it NGC~7743}. ---
The bright Seyfert~2 nucleus of NGC 7743 is surrounded by clumpy dust, 
clearly seen both in our residual image and in an archival $V$-band image.
The galaxy centroid is badly off-centered on the NIC2 image and lies in 
the upper right quadrant. Thus, only the region within $r$ = 6\asec\ was used 
for the analysis presented here.

%\clearpage
%
%\vskip 0.75truein
%\centerline{\bf{References}}
%\medskip

%\noindent

%FIGURE CAPTIONS
\clearpage
\centerline{FIGURE CAPTIONS}
\bigskip

{\it Fig. 1{\it a, b}. ---}
({\it a}) NGC 221 and NGC 404.  ({\it b}) NGC 474 and NGC 524.
{\it Top:} The observed NICMOS F160W image ({\it left}) and residual image
({\it right}). Positive values are dark, and negative values are white.
The images are 10\asec\ $\times$ 10\asec\ for NIC2 data
and 25\asec\ $\times$ 25\asec\ for NIC3 data, centered on the
galaxy, with North oriented up and East to the left.  {\it Bottom:} Results
of the isophotal analysis.  The panels plot the variation of surface
brightness $\mu_H$, ellipticity $\epsilon$, P.A., and the shape parameter
$B_4$ along the semi-major axis.  The best-fitting Nuker function, derived
from the 2-D model, is overplotted for $\mu_H$ with ({\it solid line}) and 
without ({\it dotted line}) a central point source.  In the bottom three 
panels, the vertical {\it dashed line}\ indicates the region interior to 
which the points are strongly affected by the PSF.
 
\bigskip
{\it Fig. 1{\it c, d}. ---}
({\it c}) NGC 821 and NGC 1052.  ({\it d}) NGC 2685 and NGC 3115.  As in
Fig.~1{\it a}, 1{\it b}. {\it Top:} The observed NICMOS F160W image 
({\it left}) and residual image ({\it right}).  {\it Bottom:} Results of the 
isophotal analysis.
 
\bigskip
{\it Fig. 1{\it e, f}. ---}
({\it e}) NGC 3379 and NGC 3384.  ({\it f}) NGC 3593 and NGC 3900.  As in
Fig.~1{\it a}, 1{\it b}. {\it Top:} The observed NICMOS F160W image  
({\it left}) and residual image ({\it right}).  {\it Bottom:} Results of the  
isophotal analysis.

\bigskip
{\it Fig. 1{\it g, h}. ---}
({\it g}) NGC 4026 and NGC 4111.  ({\it h}) NGC 4143 and NGC 4150.  As in
Fig.~1{\it a}, 1{\it b}. {\it Top:} The observed NICMOS F160W image  
({\it left}) and residual image ({\it right}).  {\it Bottom:} Results of the  
isophotal analysis.
 
\bigskip
{\it Fig. 1{\it i, j}. ---}
({\it i}) NGC 4261 and NGC 4278.  ({\it j}) NGC 4291 and NGC 4374.  As in
Fig.~1{\it a}, 1{\it b}. {\it Top:} The observed NICMOS F160W image  
({\it left}) and residual image ({\it right}).  {\it Bottom:} Results of the  
isophotal analysis.

\bigskip
{\it Fig. 1{\it k, l}. ---}
({\it k}) NGC 4406 and NGC 4417.  ({\it l}) NGC 4472 and NGC 4589.  As in
Fig.~1{\it a}, 1{\it b}. {\it Top:} The observed NICMOS F160W image  
({\it left}) and residual image ({\it right}).  {\it Bottom:} Results of the  
isophotal analysis.

\bigskip
{\it Fig. 1{\it m, n}. ---}
({\it m}) NGC 4636 and NGC 5273.  ({\it n}) NGC 5548 and NGC 5838.  As in
Fig.~1{\it a}, 1{\it b}. {\it Top:} The observed NICMOS F160W image  
({\it left}) and residual image ({\it right}).  {\it Bottom:} Results of the  
isophotal analysis.
 
\bigskip
{\it Fig. 1{\it o, p}. ---}
({\it o}) NGC 5982 and NGC 6340.  ({\it p}) NGC 7457 and NGC 7626.  As in
Fig.~1{\it a}, 1{\it b}. {\it Top:} The observed NICMOS F160W image  
({\it left}) and residual image ({\it right}).  {\it Bottom:} Results of the  
isophotal analysis.
 
\bigskip
{\it Fig. 1{\it q}. ---}
({\it q}) NGC 7743.  As in
Fig.~1{\it a}, 1{\it b}. {\it Top:} The observed NICMOS F160W image  
({\it left}) and residual image ({\it right}).  {\it Bottom:} Results of the  
isophotal analysis.

\bigskip
{\it Fig. 2. ---}
Comparison of aperture photometry derived from NICMOS and ground-based 
images.  The measurements have been made with $\sim$10\asec~diameter 
apertures.  The solid line denotes equality.  The average difference 
between the two sets of data is $\langle{\Delta m_H}\rangle\,=\langle m_H(HST)\,-\,
m_H({\rm ground})\rangle$ = $-$0.006$\pm$0.11 mag.

\bigskip
{\it Fig. 3. ---}
Dependence of the inner cusp slope $\gamma$  on ({\it a}) absolute 
magnitude of the galaxy ($M^0_{B_T}$) and ({\it b}) break radius ($r_b$). 
The distribution of $\gamma$ is shown in panel ({\it c}).  {\it Filled}\ 
symbols represent core galaxies, and {\it open }\ symbols denote power-law 
galaxies.  The galaxies from this study are plotted as {\it circles},
those from Faber et al. (1997) are shown as {\it triangles}, and those from
Rest et al. (2001) as {\it squares}.  Objects in our sample which fall in the 
``gap region'' (0.3 $< \gamma <$ 0.5) appear as {\it asterisks}, while those 
from Rest et al. (2001) are shown by {\it plus} signs. Two galaxies which 
fall in this region from Faber et al. (1997) are shown by {\it crosses}.

\bigskip
{\it Fig. 4. ---}
Comparison of the values of ({\it a}) $\gamma$, ({\it b}) $\beta$, and 
({\it c}) $r_b$ derived from NICMOS images to those derived from optical 
images. The solid line denotes equality.  Filled circles correspond to 
NGC 524, NGC 4589 and NGC 7626, which are known to have dust in the central 
regions as seen on optical images.  The central profiles of these objects 
most likely were depressed in the optical by dust extinction, leading to 
$\gamma$(opt) $\approx$ 0.

\bigskip
{\it Fig. 5. ---}
Central-parameter relations for core galaxies. The core parameters ($r_{b}$
and  $\mu_{b}$) are well correlated with the central stellar velocity
dispersion ($\sigma_0$) and, with greater scatter, with the total $B$-band
absolute magnitude ($M^0_{B_T}$).  Panels ({\it a}) and ({\it b}) plot only
data from this study ({\it circles}), whereas panels ({\it c}) and ({\it d})
include measurements from Faber et al. (1997; {\it triangles}) and 
Rest et al. (2001; {\it squares}).
The outliers NGC 404 and NGC 4636 are labeled.

\bigskip
{\it Fig. 6. ---}
Fundamental-plane relation between $\mu_{b}$ and $r_{b}$ for core galaxies.
The outliers NGC 404 and NGC 4636 are labeled.

\bigskip
{\it Fig. 7. ---}
Correlation of nuclear point-source magnitude with ({\it a}) 
extinction-corrected H$\alpha$ luminosity and ({\it b}) central stellar
velocity dispersion.  {\it Filled}\ symbols denote nuclei in core galaxies, 
{\it open} symbols represent nuclei in power-law galaxies, and 
{\it asterisks}\ mark objects with intermediate inner slopes 
(0.3 $< \gamma <$ 0.5).  Upper limits are indicated with arrows. Two objects 
with uncertain nuclear profiles appear as {\it triangles}. 

%\clearpage
 
%TABLES
%Table 1 has bounding box = 30 80 600 720
%\clearpage
\begin{figure}
\plotone{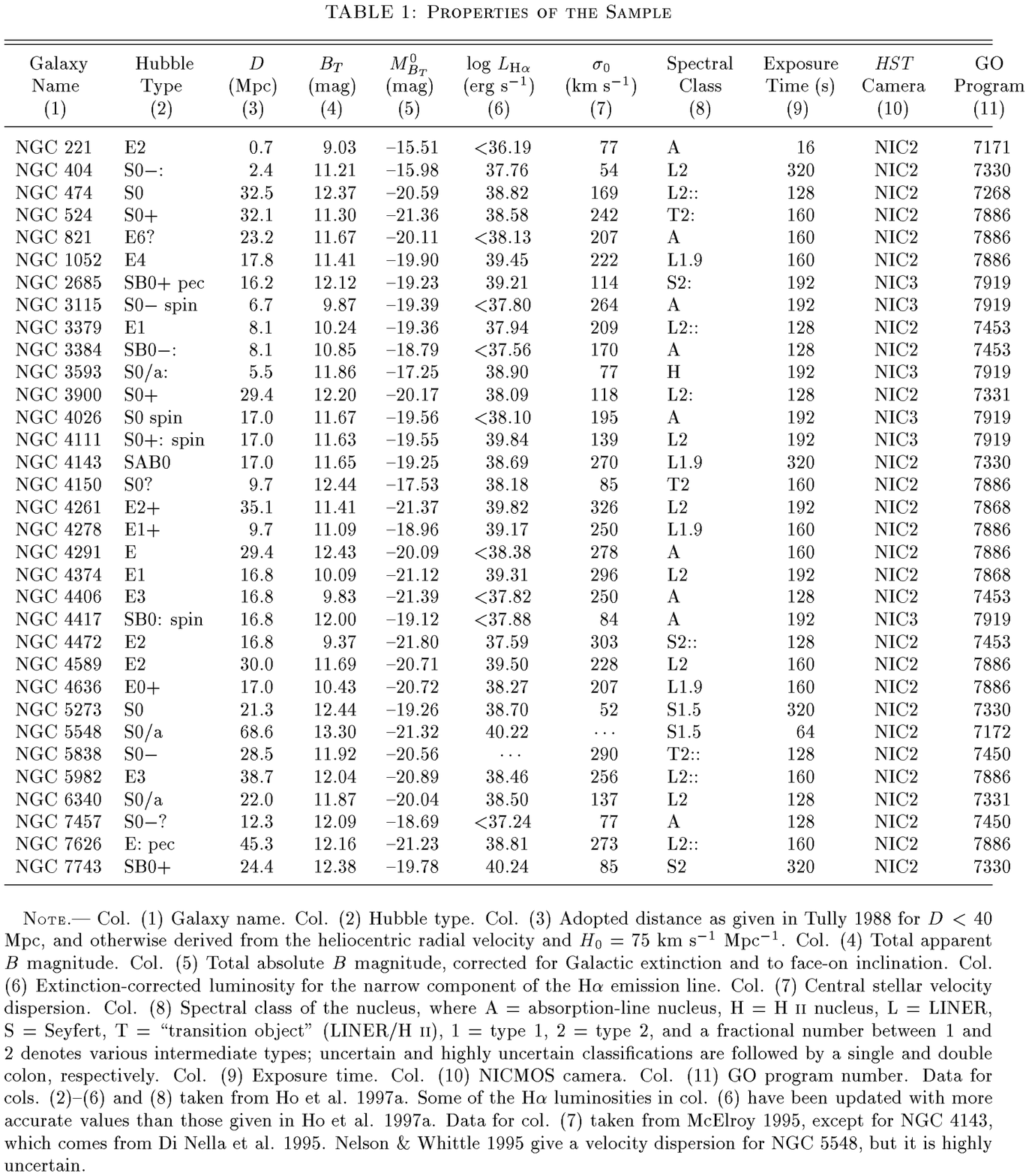}
\hskip 0.5in
\end{figure}

%Table 2 has bounding box = 10 100 580 700
%\clearpage
\begin{figure}
\plotone{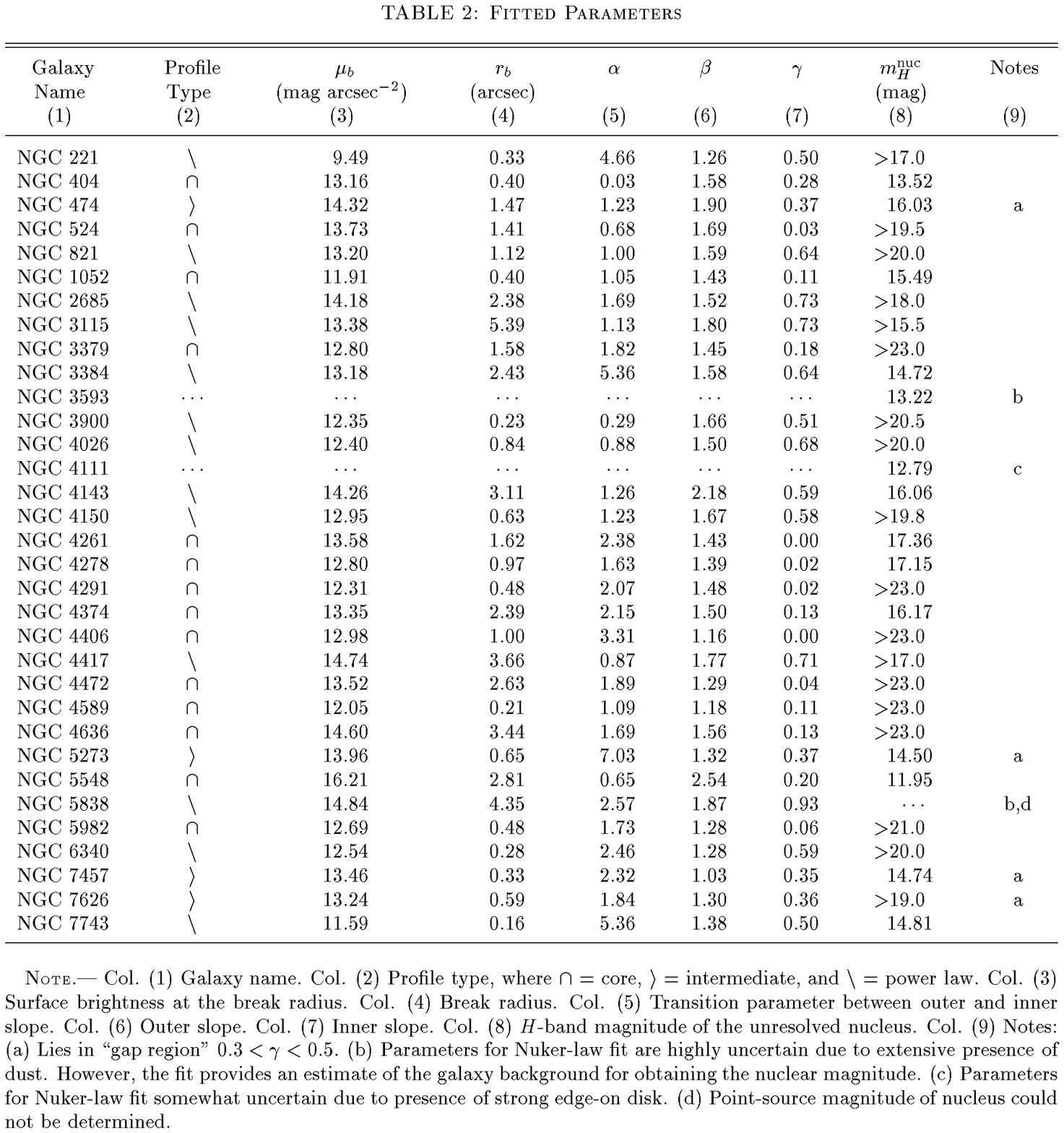}
\end{figure}

%Table 3 has bounding box = 40 260 590 540
%\clearpage
\begin{figure}
\plotone{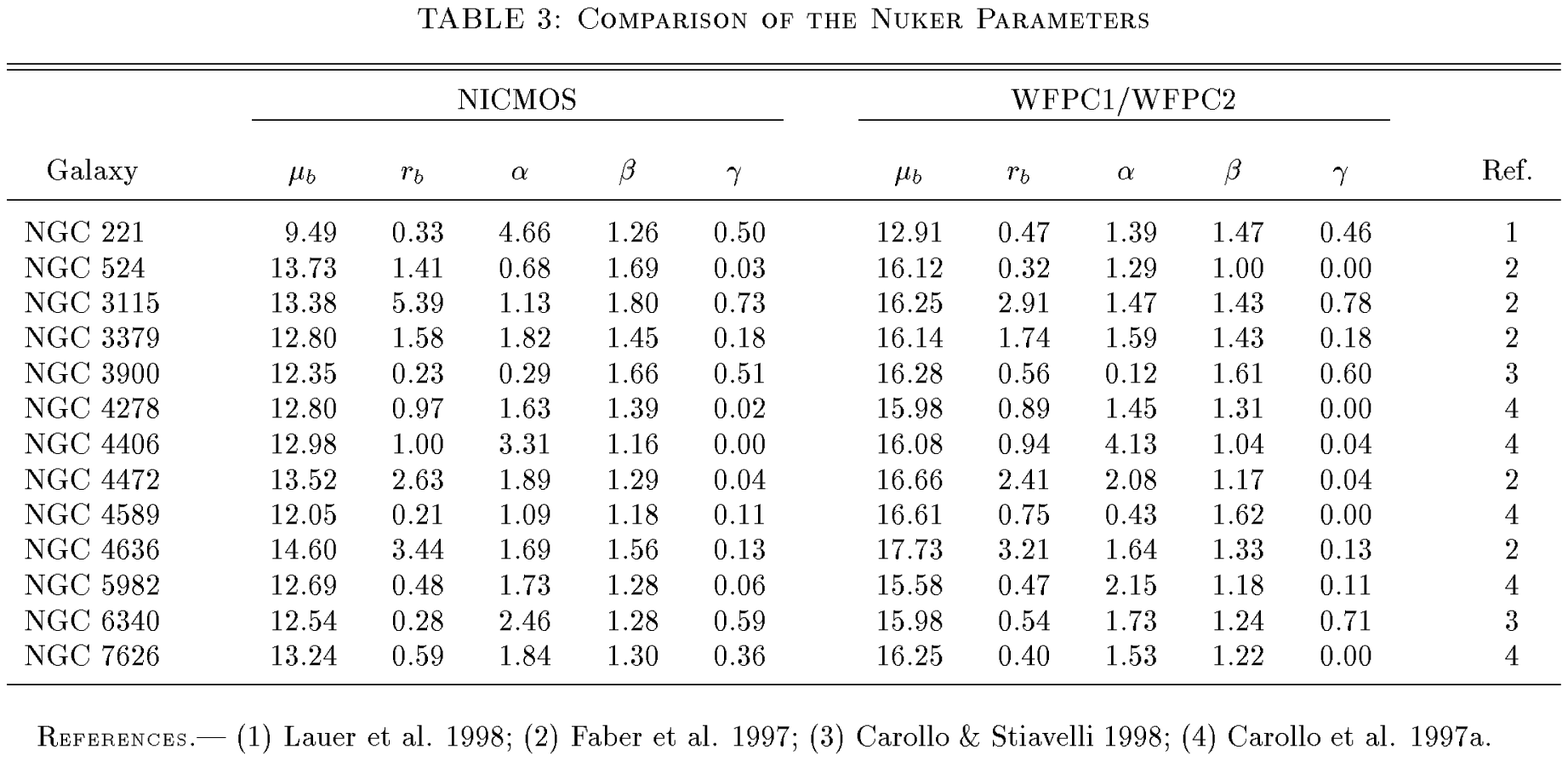}
\end{figure}

%FIGURES
\clearpage
\begin{figure}
\figurenum{2}
%\plotone{fig2.ps}
\psfig{file=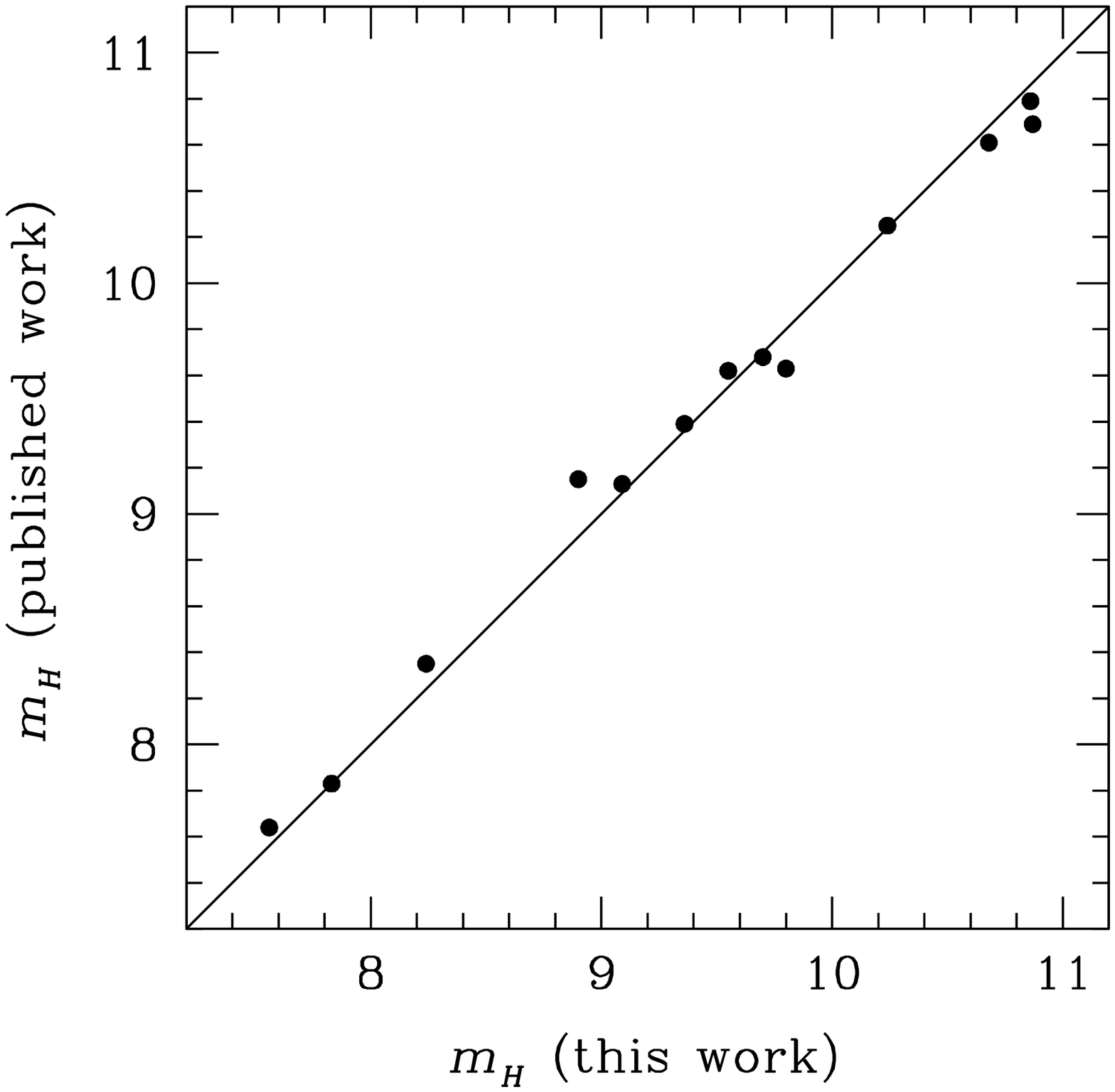,height=6.5in,angle=0}
\caption{
Comparison of aperture photometry derived from NICMOS and ground-based 
images.  The measurements have been made with $\sim$10\asec~diameter 
apertures.  The solid line denotes equality.  The average difference 
between the two sets of data is 
$\langle{\Delta m_H}\rangle\,=\langle m_H(HST)\,-\,
m_H({\rm ground})\rangle$ = $-$0.006$\pm$0.11 mag.
}
\end{figure}

\clearpage
\begin{figure}
\figurenum{3}
%\plotone{fig3.ps}
\psfig{file=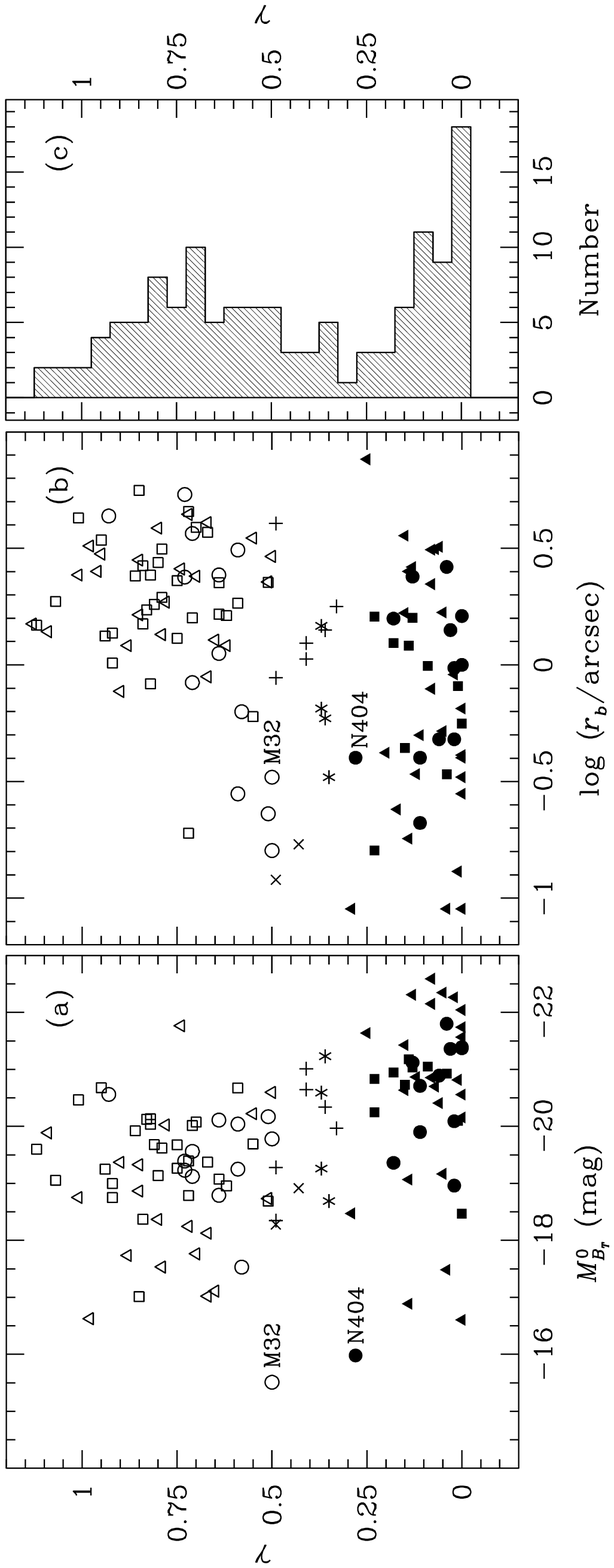,height=8.5in,angle=0}
\caption{
Dependence of the inner cusp slope $\gamma$  on ({\it a}) absolute 
magnitude of the galaxy ($M^0_{B_T}$) and ({\it b}) break radius ($r_b$). 
The distribution of $\gamma$ is shown in panel ({\it c}).  {\it Filled}\ 
symbols represent core galaxies, and {\it open }\ symbols denote power-law 
galaxies.  The galaxies from this study are plotted as {\it circles},
those from Faber et al. (1997) are shown as {\it triangles}, and those from
Rest et al. (2001) as {\it squares}.  Objects in our sample which fall in the 
``gap region'' (0.3 $< \gamma <$ 0.5) appear as {\it asterisks}, while those 
from Rest et al. (2001) are shown by {\it plus} signs. Two galaxies which 
fall in this region from Faber et al. (1997) are shown by {\it crosses}.
}
\end{figure}

\clearpage
\begin{figure}
\figurenum{4}
%\plotone{fig4.ps}
\psfig{file=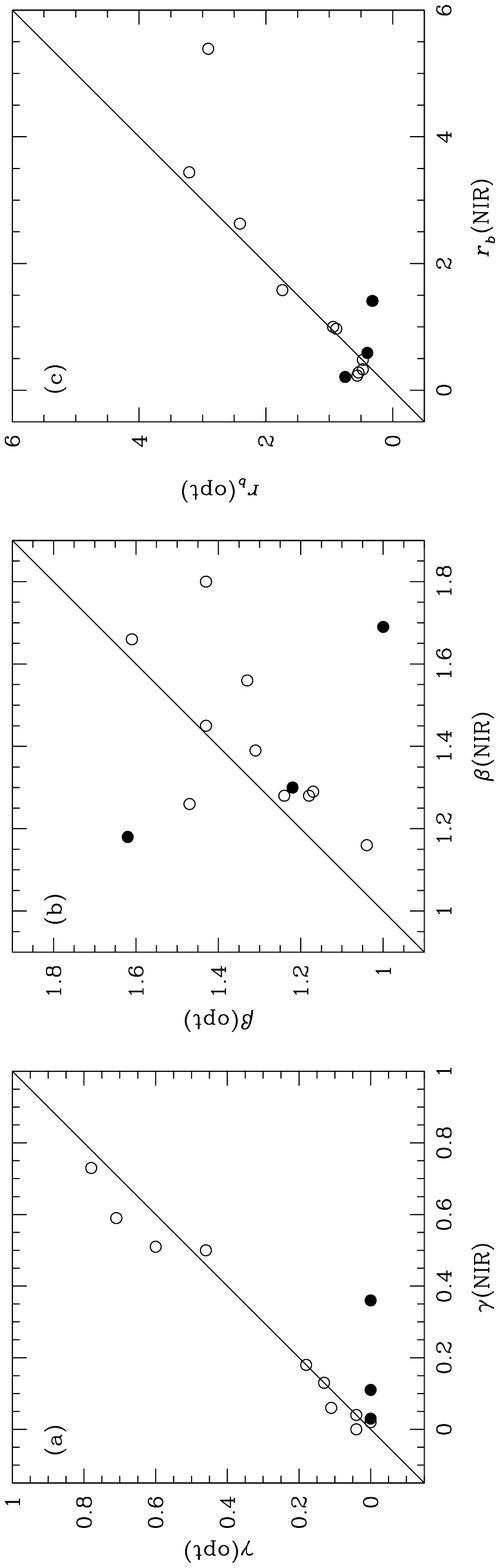,height=8.5in,angle=0}
\caption{
Comparison of the values of ({\it a}) $\gamma$, ({\it b}) $\beta$, and 
({\it c}) $r_b$ derived from NICMOS images to those derived from optical 
images. The solid line denotes equality.  Filled circles correspond to 
NGC 524, NGC 4589 and NGC 7626, which are known to have dust in the central 
regions as seen on optical images.  The central profiles of these objects 
most likely were depressed in the optical by dust extinction, leading to 
$\gamma$(opt) $\approx$ 0.
}
\end{figure}

\clearpage
\begin{figure}
\figurenum{5}
\plotone{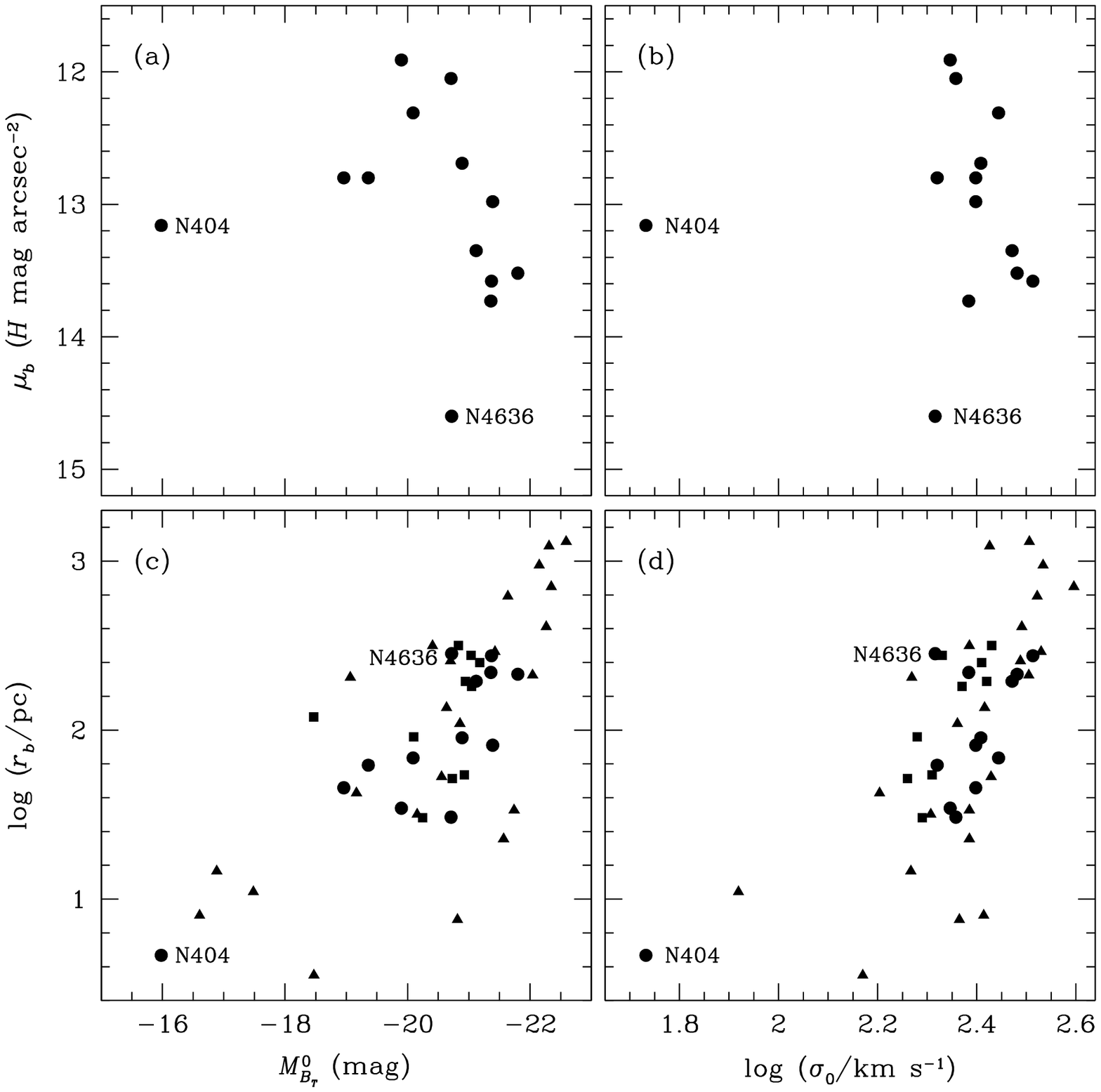}
\caption{
Central-parameter relations for core galaxies. The core parameters ($r_{b}$
and  $\mu_{b}$) are well correlated with the central stellar velocity
dispersion ($\sigma_0$) and, with greater scatter, with the total $B$-band
absolute magnitude ($M^0_{B_T}$).  Panels ({\it a}) and ({\it b}) plot only
data from this study ({\it circles}), whereas panels ({\it c}) and ({\it d})
include measurements from Faber et al. (1997; {\it triangles}) and 
Rest et al. (2001; {\it squares}).
The outliers NGC 404 and NGC 4636 are labeled.
}
\end{figure}

\clearpage
\begin{figure}
\figurenum{6}
%\plotone{fig6.ps}
\psfig{file=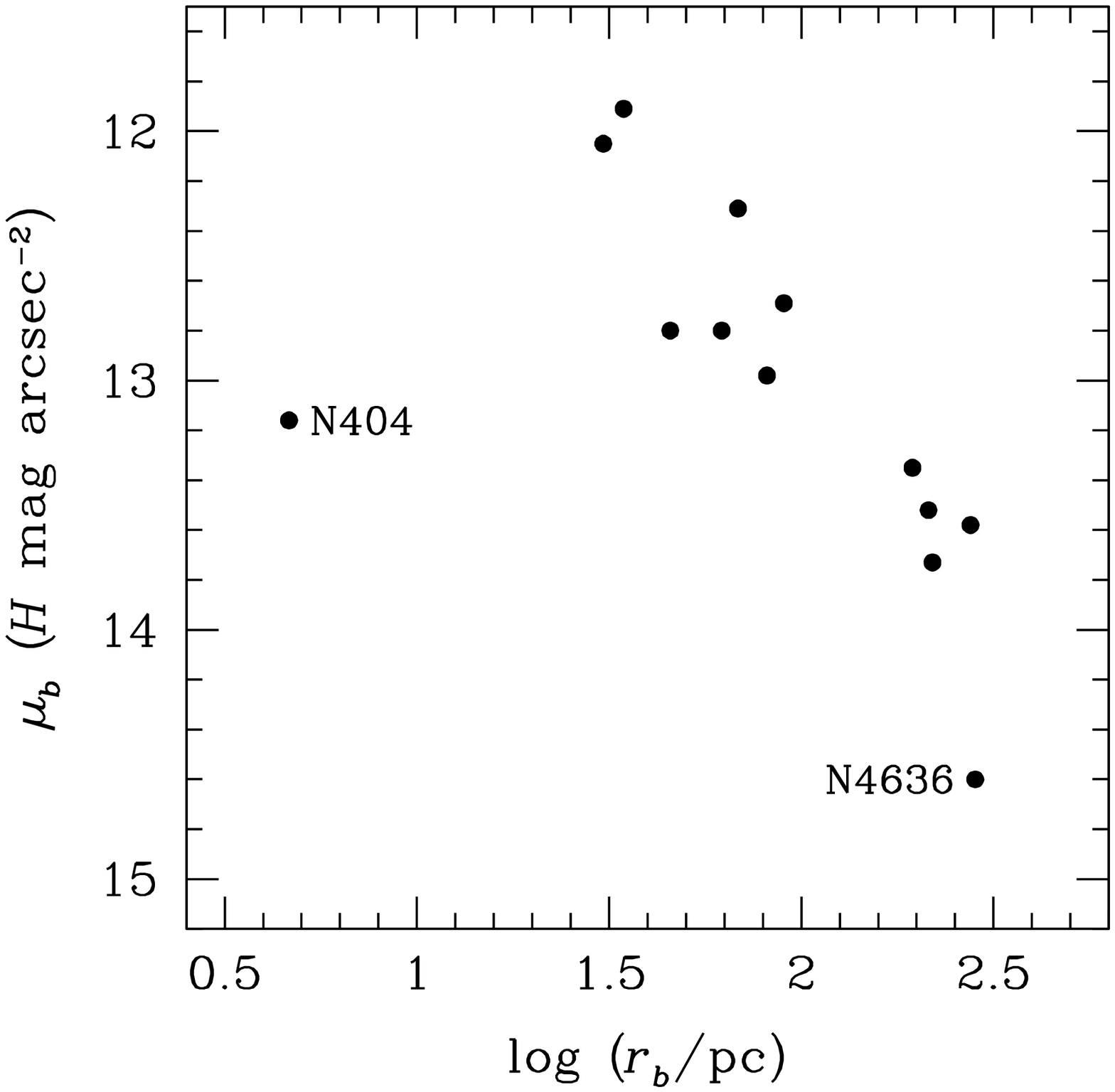,height=6.5in,angle=0}
\caption{
Fundamental-plane relation between $\mu_{b}$ and $r_{b}$ for core galaxies.
The outliers NGC 404 and NGC 4636 are labeled.
}
\end{figure}

\clearpage
\begin{figure}
\figurenum{7}
%\plotone{fig7.ps}
\psfig{file=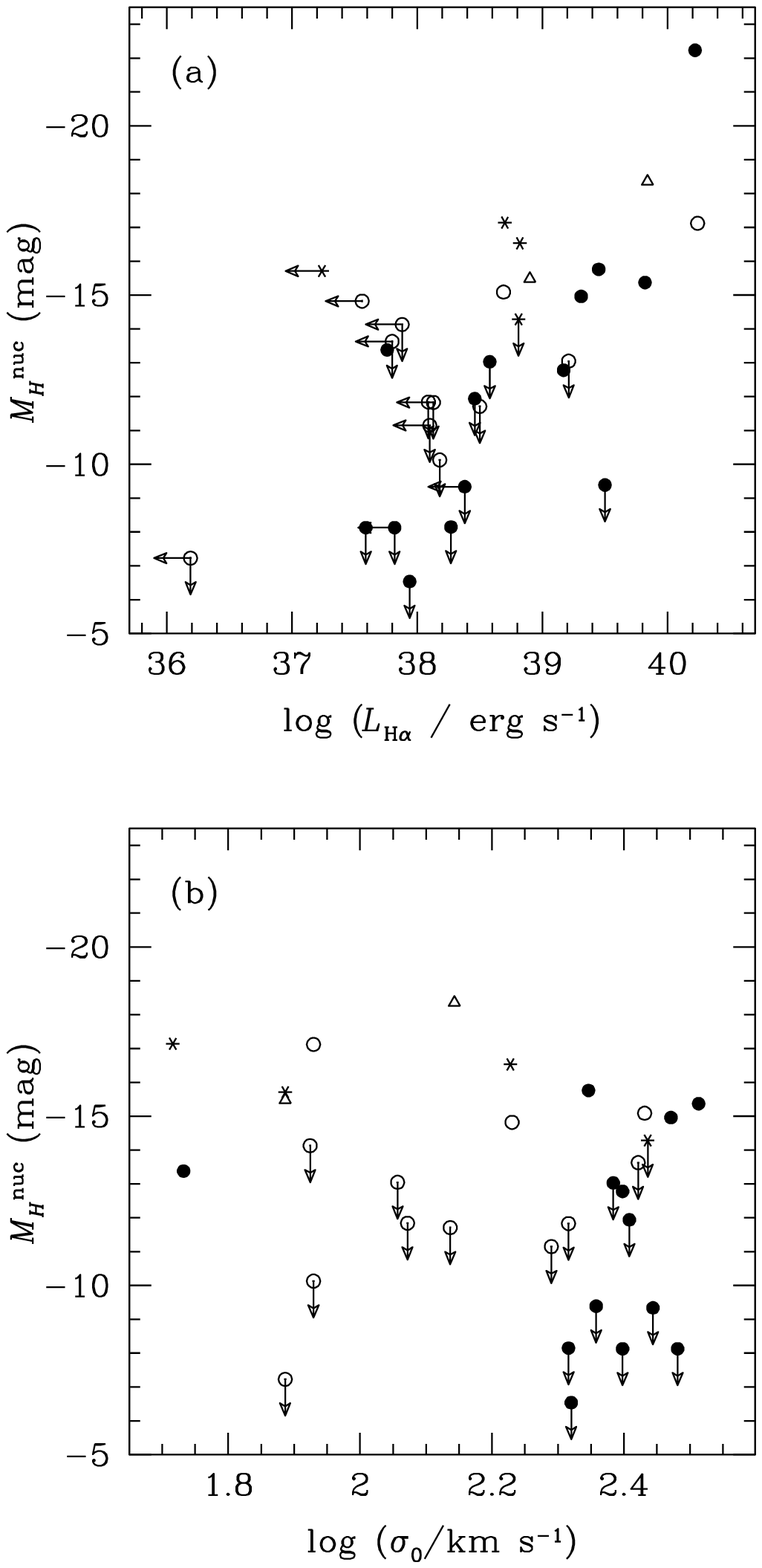,height=8.5in,angle=0}
\caption{
Correlation of nuclear point-source magnitude with ({\it a}) 
extinction-corrected H$\alpha$ luminosity and ({\it b}) central stellar
velocity dispersion.  {\it Filled}\ symbols denote nuclei in core galaxies, 
{\it open} symbols represent nuclei in power-law galaxies, and 
{\it asterisks}\ mark objects with intermediate inner slopes 
(0.3 $< \gamma <$ 0.5).  Upper limits are indicated with arrows. Two objects 
with uncertain nuclear profiles appear as {\it triangles}.
}
\end{figure}

\end{document}